\RequirePackage{fix-cm}

\documentclass[smallextended]{svjour3}
\usepackage{wrapfig,epsf}

\usepackage{amsmath}
\usepackage{amsfonts}
\usepackage{float}
\floatstyle{ruled}
\usepackage{url}
\usepackage{textcomp}
\usepackage{graphicx}
\usepackage{epstopdf}

\usepackage{algorithm}
\usepackage[noend]{algorithmic}
\usepackage{multirow}
\usepackage{flushend}

\newcommand{\nt}{\noindent}

\newcommand{\bs}[1]{\ensuremath{\boldsymbol{#1}}}
\newcommand{\mc}[1]{\ensuremath{\mathcal{#1}}}

\newcommand{\tr}[1]{trace\left( #1 \right)}
\newcommand{\tl}[1]{\tilde{#1}}
\newcommand{\proj}[1]{P^{\left(\mc{#1}\right)}}
\newcommand{\pth}[1]{\left(#1\right)}
\newcommand{\norm}[1]{\left\Vert#1\right\Vert}

\newcommand{\LR}[2]{\tl{#1}_\mc{#2}}

\newcommand{\sS}{\mc{S}}        % Set S
\newcommand{\sP}{\mc{P}}        % Set P
\newcommand{\sR}{\mc{R}}        % Set R
\newcommand{\bigO}[1]{O\left( #1 \right)}
\newcommand{\bigOO}{O}
\newcommand{\inv}[1]{\left( #1 \right)^{-1}}

\newcommand{\BlackBox}{\rule{1.5ex}{1.5ex}}

\begin{document}
%\label{firstpage}

%\renewcommand{\harvardurl}{URL: \url}
\title{Greedy Column Subset Selection for Large-scale Data Sets %\thanks{Grants or other notes
%about the article that should go on the front page should be
%placed here. General acknowledgments should be placed at the end of the article.}
}

\author{Ahmed K. Farahat  \and
        Ahmed Elgohary \and
        Ali Ghodsi \and
        Mohamed S. Kamel
}

%\authorrunning{Short form of author list} % if too long for running head

%\institute{F. Author \at
%              first address \\
%              Tel.: +123-45-678910\\
%              Fax: +123-45-678910\\
%              \email{fauthor@example.com}           %  \\
%%             \emph{Present address:} of F. Author  %  if needed
%           \and
%           S. Author \at
%              second address
%}

\institute{Ahmed K. Farahat \at
              Department of Electrical and Computer Engineering \\
              University of Waterloo \\
              Waterloo, Ontario, Canada N2L 3G1 \\
              Tel.: +1 519-888-4567\\
%              Fax: +123-45-678910\\
              \email{afarahat@uwaterloo.ca}           %  \\
%             \emph{Present address:} of F. Author  %  if needed
           \and
           Ahmed Elgohary \at
              Department of Electrical and Computer Engineering \\
              University of Waterloo \\
              Waterloo, Ontario, Canada N2L 3G1 \\
              Tel.: +1 519-888-4567\\
%              Fax: +123-45-678910\\
              \email{aelgohary@uwaterloo.ca}           %  \\
           \and
           Ali Ghodsi \at
              Department of Statistics and Actuarial Science \\
              University of Waterloo \\
              Waterloo, Ontario, Canada N2L 3G1 \\
              Tel.: +1 519-888-4567 x37316\\
%              Fax: +123-45-678910\\
              \email{aghodsib@uwaterloo.ca}
           \and
           Mohamed S. Kamel \at
              Department of Electrical and Computer Engineering \\
              University of Waterloo \\
              Waterloo, Ontario, Canada N2L 3G1 \\
              Tel.: +1 519-888-4567 x35761\\
%              Fax: +123-45-678910\\
              \email{mkamel@uwaterloo.ca}
}
\date{Received: date / Accepted: date}

\maketitle

%\begin{abstract}
%Insert your abstract here. Include keywords, PACS and mathematical
%subject classification numbers as needed.
%\keywords{First keyword \and Second keyword \and More}
%% \PACS{PACS code1 \and PACS code2 \and more}
%% \subclass{MSC code1 \and MSC code2 \and more}
%\end{abstract}

\begin{abstract}
In today's information systems, the availability of massive amounts of data necessitates the development of fast and accurate algorithms to summarize these data and represent them in a succinct format. One crucial problem in big data analytics is the selection of representative instances from large and massively-distributed data, which is formally known as the Column Subset Selection (CSS) problem. The solution to this problem enables data analysts to understand the insights of the data and explore its hidden structure. The selected instances can also be used for data preprocessing tasks such as learning a low-dimensional embedding of the data points or computing a low-rank approximation of the corresponding matrix. This paper presents a fast and accurate greedy algorithm for large-scale column subset selection. The algorithm minimizes an objective function which measures the reconstruction error of the data matrix based on the subset of selected columns. The paper first presents a centralized greedy algorithm for column subset selection which depends on a novel recursive formula for calculating the reconstruction error of the data matrix. The paper then presents a MapReduce algorithm which selects a few representative columns from a matrix whose columns are massively distributed across several commodity machines. The algorithm first learns a concise representation of all columns using random projection, and it then solves a generalized column subset selection problem at each machine in which a subset of columns are selected from the sub-matrix on that machine such that the reconstruction error of the concise representation is minimized. The paper demonstrates the effectiveness and efficiency of the proposed algorithm through an empirical evaluation on benchmark data sets.\footnote{A preliminary
version of this paper appeared as \cite{farahat13-icdm}}
\end{abstract}

% ICDM Abstract
% Given a very large data set distributed over a cluster of several nodes, this paper addresses the problem of selecting a few data instances that best represent the entire data set. The solution to this problem is of a crucial importance in the big data era as it enables data analysts to understand the insights of the data and explore its hidden structure. The selected instances can also be used for data preprocessing tasks such as learning a low-dimensional embedding of the data points or computing a low-rank approximation of the corresponding matrix. The paper first formulates the problem as the selection of a few representative columns from a matrix whose columns are massively distributed, and it then proposes a MapReduce algorithm for selecting those representatives. The algorithm first learns a concise representation of all columns using random projection, and it then solves a generalized column subset selection problem at each machine in which a subset of columns are selected from the sub-matrix on that machine such that the reconstruction error of the concise representation is minimized. The paper then demonstrates the effectiveness and efficiency of the proposed algorithm through an empirical evaluation on benchmark data sets.

\section{Introduction} \label{Sec:Intro}
Recent years have witnessed the rise of the big data era in computing and storage systems. With the great advances in information and communication technology, hundreds of petabytes of data are generated, transferred, processed and stored every day. The availability of this overwhelming amount of structured and unstructured data creates an acute need to develop fast and accurate algorithms to discover useful information that is hidden in the big data. One of the crucial problems in the big data era is the ability to represent the data and its underlying information in a succinct format.

Although different algorithms for clustering and dimension reduction can be used to summarize big data, these algorithms tend to learn representatives whose meanings are difficult to interpret. For instance, the traditional clustering algorithms such as $k$-means \cite{Jain88} tend to produce centroids which encode information about thousands of data instances. The meanings of these centroids are hard to interpret. Even clustering methods that use data instances as prototypes, such as $k$-medoid \cite{kaufman1987cmm}, learn only one representative for each cluster, which is usually not enough to capture the insights of the data instances in that cluster. In addition, using medoids as representatives implicitly assumes that the data points are distributed as clusters and that the number of those clusters are known ahead of time. This assumption is not true for many data sets. On the other hand, traditional dimension reduction algorithms such as Latent Semantic Analysis (LSA) \cite{deerwester1990ils} tend to learn a few latent concepts in the feature space. Each of these concepts is represented by a dense vector which combines thousands of features with positive and negative weights. This makes it difficult for the data analyst to understand the meaning of these concepts. Even if the goal of representative selection is to learn a low-dimensional embedding of data instances, learning dimensions whose meanings are easy to interpret allows the understanding of the results of the data mining algorithms, such as understanding the meanings of data clusters in the low-dimensional space.

The acute need to summarize big data to a format that appeals to data analysts motivates the development of different algorithms to directly select a few representative data instances and/or features. This problem can be generally formulated as the selection of a subset of columns from a data matrix, which is formally known as the Column Subset Selection (CSS) problem \cite{Frieze98-Rnd}, \cite{Drineas06-Cols} \cite{Boutsidis08-Clust} \cite{Boutsidis09a-CSS} \cite{Boutsidis11a-NearOpt}. Although many algorithms have been proposed for tackling the CSS problem, most of these algorithms focus on randomly  selecting a subset of columns with the goal of using these columns to obtain a low-rank approximation of the data matrix. In this case, these algorithms tend to select a relatively large number of columns. When the goal is to select a very few columns to be directly presented to a data analyst or indirectly used to interpret the results of other algorithms, the randomized CSS methods are not going to produce a meaningful subset of columns. One the other hand, deterministic algorithms for CSS, although more accurate, do not scale to work on big matrices with massively distributed columns.

This paper addresses the aforementioned problem by first presenting a fast and accurate greedy algorithm for column subset selection. The algorithm minimizes an objective function which measures the reconstruction error of the data matrix based on the subset of selected columns. The paper presents a novel recursive formula for calculating the reconstruction error of the data matrix, and then proposes a fast and accurate algorithm which selects the most representative columns in a greedy manner. The paper then presents a distributed column subset selection algorithm for selecting a very few columns from a big data matrix with massively distributed columns. The algorithm starts by learning a concise representation of the data matrix using random projection. %This concise representation is learned efficiently on MapReduce and then propagated to all machines.
Each machine then independently solves a generalized column subset selection problem in which a subset of columns is selected from the current sub-matrix such that the reconstruction error of the concise representation is minimized. A further selection step is then applied to the columns selected at different machines to select the required number of columns. The proposed algorithm is designed to be executed efficiently over massive amounts of data stored on a cluster of several commodity nodes. In such settings of infrastructure, ensuring the scalability and the fault tolerance of data processing jobs is not a trivial task. In order to alleviate these problems, MapReduce \cite{Dean2008} was introduced to simplify large-scale data analytics over a distributed environment of commodity machines. Currently, MapReduce %\cite{Dean2008}
(and its open source implementation Hadoop \cite{hadoop})
is considered the most successful and widely-used framework for managing big data processing jobs.
The approach proposed in this paper considers the different aspects of developing MapReduce-efficient algorithms.

%One important goal of this work, it to develop an algorithm that can be executed efficiently over massive amounts of data stored on a cluster of several nodes.
%
%The goal of this paper is to develop a Column Subset Selection (CSS) algorithm that can be executed efficiently over massive amounts of data stored on a cluster of several nodes. In such settings of infrastructure, ensuring the scalability and the fault tolerance of data processing jobs is not a trivial task. In order to alleviate these problems, a programming model, namely MapReduce \cite{Dean2008}, was introduced to  to simplify large-scale data analytics over a distributed environment of commodity machines. Currently, the MapReduce framework and its open source implementation Hadoop \cite{hadoop} are considered the most successful and widely-used framework for managing data processing jobs on distributed clusters of several nodes.

The contributions of the paper can be summarized as follows:
\begin{itemize}
  \item The paper first presents a fast and accurate algorithm for Column Subset Selection (CSS) which selects the most representative columns from a data matrix in a greedy manner. The algorithm minimizes an objective function which measures the reconstruction error of the data matrix based on the subset of selected columns.
  \item The paper presents a novel recursive formula for calculating the reconstruction error of the data matrix based on the subset of selected columns, and then uses this formula to develop a fast and accurate algorithm for greedy CSS.
  \item The paper proposes an algorithm for distributed CSS which first learns a concise representation of the data matrix and then selects columns from distributed sub-matrices that approximate this concise representation.
  \item To facilitate CSS from different sub-matrices, a fast and accurate algorithm for generalized CSS is proposed. This algorithm greedily selects a subset of columns from a source matrix which approximates the columns of a target matrix.
  \item A MapReduce-efficient algorithm is proposed for learning a concise representation using random projection. The paper also presents a MapReduce algorithm for distributed CSS which only requires two passes over the data with a very low communication overhead.
  \item Medium and large-scale experiments have been conducted on benchmark data sets in which different methods for CSS are compared.
\end{itemize}

The rest of the paper is organized as follows. Section \ref{Sec:Notations} describes the notations used throughout the paper. Section \ref{Sec:Background} gives a brief background on the CSS problem and the MapReduce framework. Section \ref{Sec:GreedyCSS} describes the centralized greedy algorithm for CSS. The proposed MapReduce algorithm for distributed CSS is described in details in Section \ref{Sec:DistCSS}. Section \ref{Sec:RelatedWork} reviews the state-of-the-art CSS methods and their applicability to distributed data. In Section \ref{Sec:Experiments}, an empirical evaluation of the proposed method is described. Finally, Section \ref{Sec:Conclusion} concludes the paper.

\section{Notations} \label{Sec:Notations}

The following notations are used throughout the paper unless otherwise indicated. Scalars are denoted by small letters (e.g., $m$, $n$), sets are denoted in script letters (e.g., $\mc{S}$, $\mc{R}$), vectors are denoted by small bold italic letters (e.g., $\bs{f}$, $\bs{g}$), and matrices are denoted by capital letters (e.g., $A$, $B$). The subscript $\pth{i}$ indicates that the variable corresponds to the $i$-th block of data in the distributed environment. In addition, the following notations are used:

%, vectors, sets, and matrices are
%shown in small, small bold italic, script, and capital letters,
%respectively.

%
%\begin{tabular} {p{1cm}p{6.5cm}}
%$\mathbb{R}^p$ & space of real $p$-dimensional
%vectors.\\
%$\mathbb{R}^{p \times q}$ & space of real $p \times q$ matrices.\\
%$| \mc{S}|$ & cardinality of set $\mc{S}$.\\
%\end{tabular}

\nt For a set $\mc{S}$:

\begin{tabular} {p{0.9cm}p{6cm}}
 $|\mc{S}|$ &   the cardinality of the set.\\
\end{tabular}

\nt For a vector $\bs{x} \in \mathbb{R}^{m}$:

\begin{tabular} {p{0.7cm}p{9cm}}
 $\bs{x}_{i}$ &  $i$-th element of $\bs{x}$.\\
$\| \bs{x} \|$ & the Euclidean norm ($\ell_2$-norm) of $\bs{x}$.\\
\end{tabular}

\nt For a matrix $A \in \mathbb{R}^{m \times n}$:

\begin{tabular} {p{0.9cm}p{9cm}}
 $A_{ij}\;\;$ & $(i,j)$-th entry of $A$.\\
 $A_{i:}$ &  $i$-th row of $A$.\\
 $A_{:j}$ & $j$-th column of $A$.\\
 $A_{:\mc{S}}$ & the sub-matrix of $A$ which consists of the set $\mc{S}$ of columns.\\
 %$A_{(i)}$ & the sub-matrix of $A$ stored at machine $i$.\\
 $A^{T}$ & the transpose of $A$. \\
 $\Vert A\Vert_F$ & the Frobenius norm of $A$: $\Vert A\Vert_F=\sqrt{\Sigma_{i,j} A^2_{ij}}$. \\
 $\tl{A}$ & a low rank approximation of $A$. \\
 $\LR{A}{S}$ & a rank-$l$ approximation of $A$ based on the set $\mc{S}$ of columns, where $| \mc{S}| = l$. \\
\end{tabular}

\section{Background} \label{Sec:Background}
This section reviews the necessary background on the Column Subset Selection (CSS) problem and the MapReduce paradigm used to develop the large-scale CSS algorithm presented in this paper.

\subsection{Column Subset Selection (CSS)} \label{Sec:CSS}

The Column Subset Selection (CSS) problem can be generally defined as the selection of the most representative columns of a data matrix \cite{Boutsidis08-Clust} \cite{Boutsidis09a-CSS} \cite{Boutsidis11a-NearOpt}. The CSS problem generalizes the problem of selecting representative data instances as well as the unsupervised feature selection problem. Both are crucial tasks, that can be directly used for data analysis or as pre-processing steps for developing fast and accurate algorithms in data mining and machine learning.

Although different criteria for column subset selection can be defined, a common criterion that has been used in much recent work measures the discrepancy between the original matrix and the approximate matrix reconstructed from the subset of selected columns \cite{Frieze98-Rnd} \cite{Drineas04-clust} \cite{Drineas2007a} \cite{Drineas06-Cols} \cite{Deshpande06b-Vol} \cite{Boutsidis08-Clust} \cite{Boutsidis09a-CSS} \cite{Boutsidis11a-NearOpt} \cite{Civril12-CSS-Sparse}. Most of the recent work either develops CSS algorithms that directly optimize this criterion or uses this criterion to assess the quality of the proposed CSS algorithms. % In the present work, the following criterion is optimized:
% The CSS problem can be formally defined as follows:
In the present work, the CSS problem is formally defined as
\begin{problem} {\bf (Column Subset Selection)} \label{Pr:FS}
    Given an $m\times n$ matrix $A$ and an integer $l$, find a subset of columns $\mc{L}$ such that $|\mc{L}| =l$ and
    \begin{displaymath}
        \mc{L}=\underset{\mc{S}}{\arg\min}\:\|A-\proj{S}A\|_{F}^{2},
    \end{displaymath} where $\proj{S}$ is an $m\times m$ projection matrix which projects the columns of $A$ onto the span of the  candidate columns $A_{:\mc{S}}$\:.
    %where $\mathbf{F}\left(\mc{S}\right)$ is the column subset selection criterion, $\mc{S}$ is the set of the indices of the candidate columns and $\mc{L}$ is the set of the indices of the selected columns.
\end{problem}

%\begin{definition} \label{Def:Criterion} {\bf (Column Subset Selection Criterion)}
%    Let $A$ be an $m\times n$ matrix and $\mc{S}$ be the set of the indices of the candidate columns, the column subset selection criterion is defined as:
%    \begin{displaymath}
%        \mathbf{F}\left(\mc{S}\right)=\|A-\proj{S}A\|_{F}^{2} \:,
%    \end{displaymath} where $\proj{S}$ is an $m\times m$ projection matrix which projects the columns of $A$ onto the span of the set $\mc{S}$ of columns.
%\end{definition}
%

The criterion $\mathbf{F}\left(\mc{S}\right)=\|A-\proj{S}A\|_{F}^{2} $ represents the sum of squared
errors between the original data matrix $A$ and its rank-$l$
column-based approximation (where $l=|\mc{S}|$),
\begin{equation}\label{Eq:LowRank}
\LR{A}{S}=\proj{S}A \:.
\end{equation}

In other words, the criterion $\mathbf{F}\left(\mc{S}\right)$ calculates the Frobenius norm of the residual matrix $E=A-\LR{A}{S}$. Other types of matrix norms can also be used to quantify the reconstruction error. Some of the recent work on the CSS problem \cite{Boutsidis08-Clust} \cite{Boutsidis09a-CSS} \cite{Boutsidis11a-NearOpt} derives theoretical bounds for both the Frobenius and spectral norms of the residual matrix. The present work, however, focuses on developing algorithms that minimize the Frobenius norm of the residual matrix.

The projection matrix $\proj{S}$ can be calculated as
\begin{equation}\label{Eq:Projection}
\proj{S}=A_{:\mc{S}} \inv{A_{:\mc{S}}^{T}A_{:\mc{S}}} A_{:\mc{S}}^{T} \:,
\end{equation} where $A_{:\mc{S}}$ is the sub-matrix of $A$ which consists of the columns corresponding to $\mc{S}$. %It should be noted that if $\mc{S}$ is known, the term $\inv{A_{:\mc{S}}^{T}A_{:\mc{S}}} A_{:\mc{S}}^{T}A$ is the closed-form solution of least-squares problem $T^{*}=\underset{T}{\arg\min}\left\Vert A-A_{:\mc S}T\right\Vert _{F}^{2}$.

It should be noted that if the subset of columns $\mc{S}$ is known, the projection matrix $\proj{S}$ can be derived as follows. The columns of the data matrix $A$ can be approximated as linear combinations of the subset of columns $\mc{S}$:
\begin{equation*}
   \LR{A}{S} = A_{:\mc{S}}T \:,
\end{equation*}
where $T$ is an $l\times n$ matrix of coefficients which can be found by solving the following optimization problem.
\begin{equation*}
  T^{*}=\underset{T}{\arg\min}\left\Vert A-A_{:\mc S}T\right\Vert _{F}^{2}.
\end{equation*} This is a least-squares problem whose closed-form solution is
$T^{*} = \inv{A_{:\mc{S}}^{T}A_{:\mc{S}}} A_{:\mc{S}}^{T}A \:.$ Substituting with $T^{*}$ in $\LR{A}{S}$ gives
\begin{equation*}
  \LR{A}{S} = A_{:\mc{S}}T = A_{:\mc{S}}\inv{A_{:\mc{S}}^{T}A_{:\mc{S}}} A_{:\mc{S}}^{T}A = \proj{S}A \:.
\end{equation*}

The set of selected columns (i.e., data instances or features) can be directly presented to a data analyst to learn about the insights of the data, or they can be used to preprocess the data for further analysis. For instance, the selected columns can be used to obtain a low-dimensional representation of all columns into the subspace of selected ones. This representation can be calculated as follows.
\begin{enumerate}
  \item Calculate an orthonormal basis $Q$ for the selected columns,
      \begin{equation*}
         Q = orth\left( A_{:\mc{S}} \right) \:,
      \end{equation*}where $orth\left( .\right)$ is a function that orthogonalizes the columns of its input matrix and $Q$ is an $m\times l$ orthogonal matrix whose columns span the range of $A_{:\mc{S}}$ . The matrix $Q$ can be obtained by applying an orthogonalization algorithm such as the Gram-Schmidt algorithm to the columns of $A_{:\mc{S}}$, or by calculating the Singular Value Decomposition (SVD) or the QR decomposition of $A_{:\mc{S}}$ \cite{Golub1996}.
  \item Embed all columns of $A$ into the subspace of $Q$,
    \begin{equation}\label{Eq:CSS_LowDim}
        W=Q^TA \:,
    \end{equation}where $W$ is an $l\times n$ matrix whose columns represent an embedding of all columns into the subspace of selected ones.
\end{enumerate}

The selected columns can also be used to calculate a column-based low-rank approximation of $A$ \cite{Drineas06-Cols}. Given a subset $\mc{S}$ of columns with $|\sS|=l$, a rank-$l$ approximation of the data matrix $A$ can be calculated as:
\begin{equation}
\LR{A}{S}=\proj{S}A=A_{:\mc{S}} \inv{A_{:\mc{S}}^{T}A_{:\mc{S}}} A_{:\mc{S}}^{T}A \:.
\end{equation}

In order to calculate a rank-$k$ approximation of the data matrix $A$ where $k\le l$, the following procedure suggested by Boutsidis et al. \cite{Boutsidis11a-NearOpt} can be used.
\begin{enumerate}
    \item Calculate an orthonormal basis $Q$ for the columns of $A_{:\mc{S}}$ and embed all columns of $A$ into the subspace of $Q$:
    \begin{equation*}
        \begin{split}
          &Q = orth\left( A_{:\mc{S}} \right) \:, \\
          &\quad W = Q^{T}A \:,
        \end{split}
    \end{equation*}where $Q$ is an $m\times l$ orthogonal matrix whose columns span the range of $A_{:\mc{S}}$ and $W$ is an $l\times n$ matrix whose columns represent an embedding of all columns into the subspace of selected ones.
    \item Calculate the best rank-$k$ approximation of the embedded columns using Singular Value Decomposition (SVD):
        \begin{equation*}
          \tl{W}_{k} = U^{\pth{W}}_k \Sigma^{\pth{W}}_k {V^{\pth{W} T}_k} \:,
        \end{equation*}where $U^{\pth{W}}_k$ and $V^{\pth{W}}_k$ are $l\times k$ and $n\times k$ matrices whose columns represent the leading $k$ left and right singular vectors of $W$ respectively, $\Sigma^{\pth{W}}_k$ is a $k\times k$ matrix whose diagonal elements are the leading $k$ singular values of $W$, and $\tl{W}_{k}$ is the best rank-$k$ approximation of $W$.
    \item Calculate the column-based rank-$k$ approximation of $A$ as:
        \begin{equation*}
          \tl{A}_{\mc{S}, k} = Q\tl{W}_{k} \:,
        \end{equation*}where $\tl{A}_{\mc{S}, k}$ is a rank-$k$ approximation of $A$ based on the set $\mc{S}$ of columns.
\end{enumerate}
This procedure results in a rank-$k$ approximation of $A$ within the column space of $A_{:\mc{S}}$ that achieves the minimum reconstruction error in terms of Frobenius norm \cite{Boutsidis11a-NearOpt}:
\begin{equation*}
    T^*=\underset{T,\:rank\left(T\right)=k}{\arg\min}\left\Vert A-A_{:\mc{S}}T\right\Vert _{F}^{2} \:.
\end{equation*}

Moreover, the leading singular values and vectors of the low-dimensional embedding $W$ can be used to approximate those of the data matrix as follows:
\begin{equation} \label{Eq:ApproxSVD}
    \tl{U}^{\pth{A}}_{k}=QU^{\pth{W}}_{k},\qquad \tl{\Sigma}^{\pth{A}}_{k}=\Sigma^{\pth{W}}_{k},\qquad\tl{V}^{\pth{A}}_{k}=V^{\pth{W}}_{k}
\end{equation} where $\tl{U}^{\pth{A}}_k$ and $\tl{V}^{\pth{A}}_k$ are $l\times k$ and $n\times k$ matrices whose columns approximate the leading $k$ left and right singular vectors of $A$ respectively, and $\tl{\Sigma}^{\pth{A}}_k$ is a $k\times k$ matrix whose diagonal elements approximate the leading $k$ singular values of $A$.

%In Section \ref{Sec:CSS_Recursive}, a recursive formula for the selection criterion is presented. This formula allows the development of an efficient algorithm to greedily minimize $\mathbf{F}\left(\mc{S}\right)$. The greedy algorithm is presented in Section \ref{Sec:CSS_Greedy}.

\subsection{MapReduce Paradigm} \label{Sec:MapReduce}
MapReduce \cite{Dean2008} was presented as a programming model to simplify large-scale data analytics over a distributed environment of commodity machines. The rationale behind MapReduce is to impose a set of constraints on data access at each individual machine and communication between different machines to ensure both the scalability and fault-tolerance of the analytical tasks. Currently, MapReduce has been successfully used for scaling various data analysis tasks such as regression \cite{robustRegression}, feature selection \cite{signhLogistc}, graph mining \cite{hadi2008,ashraf}, and most recently kernel $k$-means clustering \cite{elgoharyCoRR13}.

A MapReduce job is executed in two phases of user-defined data transformation functions, namely, \textit{map} and \textit{reduce} phases. The input data is split into physical blocks distributed among the nodes. Each block is viewed as a list of \textit{key-value} pairs.
In the first phase, the \textit{key-value} pairs of each input block $b$ are processed by a single \textit{map} function running independently on the node where the block $b$ is stored. The \textit{key-value} pairs are provided one-by-one to the \textit{map} function. The output of the \textit{map} function is another set of intermediate \textit{key-value} pairs. The values associated with the same \textit{key} across all nodes are grouped together and provided as an input to the \textit{reduce} function in the second phase. Different groups of values are processed in parallel on different machines. The output of each \textit{reduce} function is a third set of \textit{key-value} pairs and collectively considered the output of the job. It is important to note that the set of the intermediate \textit{key-value} pairs is moved across the network between the nodes which incurs significant additional execution time when much data are to be moved. For complex analytical tasks, multiple jobs are typically chained together \cite{elsayed} and/or many rounds of the same job are executed on the input data set \cite{fastclustering}.

In addition to the programming model constraints, Karloff et al. \cite{mrc} defined a set of computational constraints that ensure the scalability and the efficiency of MapReduce-based analytical tasks. These computational constraints limit the used memory size at each machine, the output size of both the \textit{map} and \textit{reduce} functions and the number of rounds used to complete a certain tasks.

The MapReduce algorithms presented in this paper adhere to both the programming model constraints and the computational constraints. The proposed algorithm aims also at minimizing the overall running time of the distributed column subset selection task to facilitate interactive data analytics.

\section{Greedy Column Subset Selection} \label{Sec:GreedyCSS}
The column subset selection criterion presented in Section \ref{Sec:CSS} measures the reconstruction error of a data matrix based on the subset of selected columns. The minimization of this criterion is a combinatorial optimization problem whose optimal solution can be obtained in $O\pth{n^lmnl}$ \cite{Boutsidis09a-CSS}. This section describes a deterministic greedy algorithm for optimizing this criterion, which extends the greedy method for unsupervised feature selection recently proposed by Farahat et al. \cite{farahat11-icdm} \cite{farahat12}. First, a recursive formula for the CSS criterion is presented and then the greedy CSS algorithm is described in details.

%In this section, a recursive formula for the CSS criterion is presented. This formula allows the development of an efficient greedy algorithm that approximates the optimal solution of the column subset selection problem.

\subsection{Recursive Selection Criterion}\label{Sec:CSS_Recursive}
The recursive formula of the CSS criterion is based on a recursive formula for the projection matrix $P^{(\mc{S})}$ which can be derived as follows.

%\begin{definition} \label{Def:Criterion} {\bf (Column Subset Selection Criterion)}
%    Let $A$ be an $m\times n$ matrix. The column subset selection criterion is defined as:
%    \begin{displaymath}
%        \mathbf{F}\left(\mc{S}\right)=\|A-\proj{S}A\|_{F}^{2}
%    \end{displaymath} where $\mc{S}$ is the set of the indices of selected columns,
%    and $\proj{S}$ is an $m\times m$ projection matrix which
%    projects the columns of $A$ onto the span of the set $\mc{S}$ of
%    columns.
%\end{definition}

\begin{lemma}  \label{Lm:Proj}
    Given a set of columns $\mc{S}$. For any $\mc{P} \subset \mc{S}$,
        \begin{displaymath}
            \proj{S}=P^{\left(\mc{P}\right)}+R^{\left(\mc{R}\right)} \:,
        \end{displaymath}
    where $R^{\left(\mc{R}\right)}$ is a projection matrix which
    projects the columns of $E = A - P^{\left(\mc{P}\right)} A$
    onto the span of the subset $\mc{R} = \mc{S}\setminus \mc{P}$ of
    columns,
        \begin{displaymath}
            R^{\left(\mc{R}\right)} =
            E_{:\mc{R}} \inv{E_{:\mc{R}}^{T}E_{:\mc{R}}} E_{:\mc{R}}^{T} \:.
        \end{displaymath}
\end{lemma}

\proof Define a matrix $B=A_{:\mc{S}}^{T}A_{:\mc{S}}$ which
represents the inner-product over the columns of the sub-matrix
$A_{:\mc{S}}$. The projection matrix $\proj{S}$ can
be written as:
    \begin{equation}\label{Eq:Projection2}
        \proj{S}=A_{:\mc{S}}B^{-1}A_{:\mc{S}}^{T} \:.
    \end{equation}

Without loss of generality, the columns and rows of $A_{:\mc{S}}$
and $B$ in Equation (\ref{Eq:Projection2}) can be rearranged such that
the first sets of rows and columns correspond to $\mc{P}$:
    \begin{displaymath}
        A_{:\mc{S}}=\left[\begin{array}{cc} A_{:\mc{P}} &
        A_{:\mc{R}}\end{array}\right], \quad B=\left[\begin{array}{cc}
        B_{\mc{P}\mc{P}} & B_{\mc{P}\mc{R}}\\
        B_{\mc{P}\mc{R}}^{T} &
        B_{\mc{R}\mc{R}}\end{array}\right] \:,
    \end{displaymath}
where
$B_{\mc{P}\mc{P}}=A_{:\mc{P}}^{T}A_{:\mc{P}}$,
$B_{\mc{P}\mc{R}}=A_{:\mc{P}}^{T}A_{:\mc{R}}$
and $B_{\mc{R}\mc{R}}=A_{:\mc{R}}^{T}A_{:\mc{R}}$.

Let
$S=B_{\mc{R}\mc{R}}-B_{\mc{P}\mc{R}}^{T}B_{\mc{P}\mc{P}}^{-1}B_{\mc{P}\mc{R}}$
be the Schur complement \cite{lütkepohl1996handbook} of
$B_{\mc{P}\mc{P}}$ in $B$. Using the block-wise inversion formula
\cite{lütkepohl1996handbook}, $B^{-1}$ can be calculated as:
\begin{displaymath}
B^{-1}= \left[\begin{array}{cc}
        B_{\mc{P}\mc{P}}^{-1}+B_{\mc{P}\mc{P}}^{-1}B_{\mc{P}\mc{R}}S^{-1}B_{\mc{P}\mc{R}}^{T}B_{\mc{P}\mc{P}}^{-1} & -B_{\mc{P}\mc{P}}^{-1}B_{\mc{P}\mc{R}}S^{-1}\\
        -S^{-1}B_{\mc{P}\mc{R}}^{T}B_{\mc{P}\mc{P}}^{-1}
        & S^{-1}\end{array}\right]
\end{displaymath}

Substitute with $A_{:\mc{S}}$ and $B^{-1}$ in Equation
(\ref{Eq:Projection2}):
\begin{displaymath}
        \begin{split}
            \proj{S}=&\left[\begin{array}{cc}
            A_{:\mc{P}} &
            A_{:\mc{R}}\end{array}\right]
            \left[\begin{array}{cc}
            B_{\mc{P}\mc{P}}^{-1}+B_{\mc{P}\mc{P}}^{-1}B_{\mc{P}\mc{R}}S^{-1}B_{\mc{P}\mc{R}}^{T}B_{\mc{P}\mc{P}}^{-1} & -B_{\mc{P}\mc{P}}^{-1}B_{\mc{P}\mc{R}}S^{-1}\\
            -S^{-1}B_{\mc{P}\mc{R}}^{T}B_{\mc{P}\mc{P}}^{-1}
            & S^{-1}\end{array}\right] \left[\begin{array}{c}
            A_{:\mc{P}}^{T}\\ \\
            A_{:\mc{R}}^{T}\end{array}\right] \\
            =&A_{:\mc{P}} B_{\mc{P}\mc{P}}^{-1} A_{:\mc{P}}^{T}+ A_{:\mc{P}}B_{\mc{P}\mc{P}}^{-1}B_{\mc{P}\mc{R}}S^{-1}B_{\mc{P}\mc{R}}^{T}B_{\mc{P}\mc{P}}^{-1}A_{:\mc{P}}^{T} -A_{:\mc{P}}B_{\mc{P}\mc{P}}^{-1}B_{\mc{P}\mc{R}}S^{-1}A_{:\mc{R}}^{T} \\
            &-A_{:\mc{R}}S^{-1}B_{\mc{P}\mc{R}}^{T}B_{\mc{P}\mc{P}}^{-1}A_{:\mc{P}}^{T} + A_{:\mc{R}} S^{-1} A_{:\mc{R}}^{T} \:.
        \end{split}
\end{displaymath}

Take out $A_{:\mc{P}}B_{\mc{P}\mc{P}}^{-1}B_{\mc{P}\mc{R}}S^{-1}$ as a common factor from the $2^{nd}$ and $3^{rd}$ terms, and
$A_{:\mc{R}}S^{-1}$ from the $4^{th}$ and $5^{th}$ terms:
\begin{displaymath}
        \begin{split}
            \proj{S}=&A_{:\mc{P}} B_{\mc{P}\mc{P}}^{-1} A_{:\mc{P}}^{T}
            - A_{:\mc{P}}B_{\mc{P}\mc{P}}^{-1}B_{\mc{P}\mc{R}}S^{-1}
            \left( A_{:\mc{R}}^{T} - B_{\mc{P}\mc{R}}^{T}B_{\mc{P}\mc{P}}^{-1}A_{:\mc{P}}^{T} \right)\\
            &+A_{:\mc{R}}S^{-1} \left(A_{:\mc{R}}^{T} - B_{\mc{P}\mc{R}}^{T}B_{\mc{P}\mc{P}}^{-1}A_{:\mc{P}}^{T} \right) \:.
        \end{split}
\end{displaymath}

Take out $S^{-1} \left( A_{:\mc{R}}^{T} - B_{\mc{P}\mc{R}}^{T}B_{\mc{P}\mc{P}}^{-1}A_{:\mc{P}}^{T} \right)$
as a common factor from the $2^{nd}$ and $3^{rd}$ terms:
%The right-hand side can be simplified to:
    \begin{equation} \label{Eq:ProjP2}
        \begin{split}
        \proj{S}=A_{:\mc{P}}B_{\mc{P}\mc{P}}^{-1}A_{:\mc{P}}^{T}
        +\left(A_{:\mc{R}}-A_{:\mc{P}}B_{\mc{P}\mc{P}}^{-1}B_{\mc{P}\mc{R}}\right)S^{-1}\left(A_{:\mc{R}}^{T}-B_{\mc{P}\mc{R}}^{T}B_{\mc{P}\mc{P}}^{-1}A_{:\mc{P}}^{T}\right) \:.
        %\proj{S}&=\frac{1}{B_{pp}}A_{:l}A_{:l}^{T}\\
        %&+\left(A_{:\mc{R}}-\frac{1}{B_{pp}}A_{:l}B_{l\mc{R}}\right)S^{-1}\left(A_{:\mc{R}}-\frac{1}{B_{pp}}A_{:l}B_{l\mc{R}}\right)^{T}\\
        \end{split}
    \end{equation}

The first term of Equation (\ref{Eq:ProjP2}) is the projection matrix
which projects the columns of $A$ onto the span of the subset
$\mc{P}$ of columns:
$P^{\left(\mc{P}\right)}=A_{:\mc{P}}\pth{A_{:\mc{P}}^{T}A_{:\mc{P}}}^{-1}A_{:\mc{P}}^{T}=A_{:\mc{P}}B_{\mc{P}\mc{P}}^{-1}A_{:\mc{P}}^{T}$.
The second term can be simplified as follows. Let $E$ be an $m
\times n$ residual matrix which is calculated as:
$E=A-P^{\left(\mc{P}\right)}A$. The sub-matrix $E_{:\mc{R}}$ can be expressed as:
\begin{displaymath}
   E_{:\mc{R}}=A_{:\mc{R}}-P^{\left(\mc{P}\right)}A_{:\mc{R}} = A_{:\mc{R}}-A_{:\mc{P}}\pth{A_{:\mc{P}}^{T}A_{:\mc{P}}}^{-1}A_{:\mc{P}}^{T}A_{:\mc{R}}=A_{:\mc{R}}-A_{:\mc{P}}B_{\mc{P}\mc{P}}^{-1}B_{\mc{P}\mc{R}} \:.
\end{displaymath}
Since projection matrices are idempotent, then $P^{\left(\mc{P}\right)}P^{\left(\mc{P}\right)}=P^{\left(\mc{P}\right)}$ and the inner-product $E_{:\mc{R}}^{T}E_{:\mc{R}}$ can be expressed as:
\begin{displaymath}
    \begin{split}
       E_{:\mc{R}}^{T}E_{:\mc{R}} =& \pth{A_{:\mc{R}}-P^{\left(\mc{P}\right)}A_{:\mc{R}}}^T \pth{A_{:\mc{R}}-P^{\left(\mc{P}\right)}A_{:\mc{R}}} \\
       =&A_{:\mc{R}}^TA_{:\mc{R}}-A_{:\mc{R}}^TP^{\left(\mc{P}\right)}A_{:\mc{R}}-A_{:\mc{R}}^TP^{\left(\mc{P}\right)}A_{:\mc{R}}+A_{:\mc{R}}^TP^{\left(\mc{P}\right)}P^{\left(\mc{P}\right)}A_{:\mc{R}}\\
       =&A_{:\mc{R}}^TA_{:\mc{R}}-A_{:\mc{R}}^TP^{\left(\mc{P}\right)}A_{:\mc{R}}\:.
    \end{split}
\end{displaymath}
Substituting with $P^{\left(\mc{P}\right)}=A_{:\mc{P}}\pth{A_{:\mc{P}}^{T}A_{:\mc{P}}}^{-1}A_{:\mc{P}}^T$ gives
\begin{displaymath}
    \begin{split}
       E_{:\mc{R}}^{T}E_{:\mc{R}}    =A_{:\mc{R}}^TA_{:\mc{R}}-A_{:\mc{R}}^TA_{:\mc{P}}\pth{A_{:\mc{P}}^{T}A_{:\mc{P}}}^{-1}A_{:\mc{P}}^TA_{:\mc{R}}= B_{\mc{R}\mc{R}}-B_{\mc{P}\mc{R}}^{T}B_{\mc{P}\mc{P}}^{-1}B_{\mc{P}\mc{R}} = S \:.
    \end{split}
\end{displaymath}
%It can be shown that $E_{:\mc{R}}=A_{:\mc{R}}-A_{:\mc{P}}B_{\mc{P}\mc{P}}^{-1}B_{\mc{P}\mc{R}}$,and $S=E_{:\mc{R}}^{T}E_{:\mc{R}}$.
Substituting $\pth{A_{:\mc{P}}B_{\mc{P}\mc{P}}^{-1}A_{:\mc{P}}^{T}}$, $\pth{A_{:\mc{R}}-A_{:\mc{P}}B_{\mc{P}\mc{P}}^{-1}B_{\mc{P}\mc{R}}}$ and $S$ with $P^{\left(\mc{P}\right)}$, $E_{:\mc{R}}$ and $E_{:\mc{R}}^{T}E_{:\mc{R}}$ respectively, Equation (\ref{Eq:ProjP2}) can be expressed as:
\begin{displaymath}
        \begin{split}
        \proj{S}=\proj{P} + E_{:\mc{R}}\pth{E_{:\mc{R}}^{T}E_{:\mc{R}}}^{-1}E_{:\mc{R}}^T\:.
        \end{split}
\end{displaymath}

The second term is the projection matrix which projects the
columns of $E$ onto the span of the subset $\mc{R}$ of columns:
\begin{equation}\label{Eq:ProjR}
R^{\left(\mc{R}\right)} =
E_{:\mc{R}}\left(E_{:\mc{R}}^{T}E_{:\mc{R}}\right)^{-1}E_{:\mc{R}}^{T}\:.
\end{equation}
This proves that $\proj{S}$ can be written in terms
of $P^{\left(\mc{P}\right)}$ and $R$ as:
$\proj{S}=P^{\left(\mc{P}\right)}+R^{\left(\mc{R}\right)}$
\hfill\BlackBox

This means that projection matrix $\proj{S}$ can
be constructed in a recursive manner by first calculating the
projection matrix which projects the columns of $A$ onto the span of
the subset $\mc{P}$ of columns, and then calculating the projection
matrix which projects the columns of the residual matrix onto the
span of the remaining columns. Based on this lemma, a recursive
formula can be developed for $\LR{A}{S}$.

\begin{corollary} \label{Cl:LowRank}
    Given a matrix $A$ and a subset of columns $\mc{S}$. For any $\mc{P} \subset \mc{S}$,
    \begin{displaymath}
        \LR{A}{S}=\LR{A}{P}+\LR{E}{R} \:,
    \end{displaymath} where $E = A - P^{\left(\mc{P}\right)}A$, and $\LR{E}{R}$ is the low-rank
    approximation of $E$ based on the subset $\mc{R} = \mc{S}\setminus \mc{P}$ of
    columns.
\end{corollary}
\proof Using Lemma (\ref{Lm:Proj}), and substituting with
$\proj{S}$ in Equation (\ref{Eq:LowRank}) gives:
    \begin{equation}\label{Eq:Rec_A_1}
        \LR{A}{S} = P^{\left(\mc{P}\right)}A+E_{:\mc{R}}\left(E_{:\mc{R}}^{T}E_{:\mc{R}}\right)^{-1}E_{:\mc{R}}^{T}A \:.
    \end{equation}
The first term is the low-rank approximation of $A$ based on
$\mc{P}$: $\LR{A}{P}=P^{\left(\mc{P}\right)}A$.
The second term is equal to $\LR{E}{R}$ as
$E_{:\mc{R}}^{T}A = E_{:\mc{R}}^{T}E$. To prove that,
multiplying $E_{:\mc{R}}^{T}$ by $E = A -
P^{\left(\mc{P}\right)}A$ gives:
    \[
        E_{:\mc{R}}^{T}E=E_{:\mc{R}}^{T}A-E_{:\mc{R}}^{T}P^{\left(\mc{P}\right)}A \:.
    \]
Using
$E_{:\mc{R}}=A_{:\mc{R}}-P^{\left(\mc{P}\right)}A_{:\mc{R}}$,
the expression $E_{:\mc{R}}^{T}P^{\left(\mc{P}\right)}$
can be written as:
    \[
        E_{:\mc{R}}^{T}P^{\left(\mc{P}\right)}=A_{:\mc{R}}^{T}P^{\left(\mc{P}\right)}-A_{:\mc{R}}^{T}P^{\left(\mc{P}\right)}P^{\left(\mc{P}\right)} \:.
    \]
This is equal to $0$ as
$P^{\left(\mc{P}\right)}P^{\left(\mc{P}\right)}=P^{\left(\mc{P}\right)}$
(an idempotent matrix). This means that
$E_{:\mc{R}}^{T}A = E_{:\mc{R}}^{T}E$. Substituting
$E_{:\mc{R}}^{T}A$ with $E_{:\mc{R}}^{T}E$ in Equation
(\ref{Eq:Rec_A_1}) proves the corollary. \hfill\BlackBox

This means that the column-based low-rank approximation of $A$ based on the subset $\sS$ of columns can be calculated in a recursive manner by first calculating the low-rank approximation of $A$ based on the subset $\sP \subset \sS$, and then calculating the low-rank approximation of the residual matrix $E$ based on the remaining columns.

Based on Corollary (\ref{Cl:LowRank}), a recursive formula for the
column subset selection criterion can be developed as follows.

\begin{theorem}\label{Th:RecF}
    Given a set of columns $\mc{S}$. For any $\mc{P} \subset \mc{S}$,
    \begin{displaymath}
        \mathbf{F}\left(\mc{S}\right)=\mathbf{F}\left(\mc{P}\right)-\|\LR{E}{R}\|_{F}^{2} \:,
    \end{displaymath} where $E = A - P^{\left(\mc{P}\right)}A$, and $\LR{E}{R}$ is the low-rank
    approximation of $E$ based on the subset $\mc{R} = \mc{S}\setminus \mc{P}$ of
    columns.
\end{theorem}
\proof Using Corollary (\ref{Cl:LowRank}), the CSS criterion can be expressed as:
\begin{displaymath}
\begin{split}
\mathbf{F}\left(\mc{S}\right)&=\norm{A-\LR{A}{S}}_{F}^{2}=\norm{A-\LR{A}{P}-\LR{E}{R}}_{F}^{2}\\
&=\norm{E-\LR{E}{R}}_{F}^{2}=\norm{E-R^{\pth{\sR}}E}_{F}^{2}\:.
\end{split}
\end{displaymath}
Using the relation between the Frobenius norm and the trace
function,\footnote{$\|A \|_{F}^{2} = \tr{A^TA}$} the right-hand
side can be expressed as:
\begin{displaymath}
\begin{split}
\norm{E-R^{\pth{\sR}}E}_{F}^{2}&=\tr{\pth{E-R^{\pth{\sR}}E}^{T}\pth{E-R^{\pth{\sR}}E}}\\
&=\tr{E^{T}E-2E^{T}R^{\pth{\sR}}E+E^TR^{\pth{\sR}}R^{\pth{\sR}}E} \:.
\end{split}
\end{displaymath}
As
$R^{\left(\mc{R}\right)}R^{\left(\mc{R}\right)}=R^{\left(\mc{R}\right)}$ (an idempotent matrix), $\mathbf{F}\left(\mc{S}\right)$ can be expressed as:
%the expression $\LR{E}{R}^{T}\LR{E}{R}$
%can be written as:
%\[
%\LR{E}{R}^{T}\LR{E}{R}=E^TR^{\left(\mc{R}\right)}R^{\left(\mc{R}\right)}E=E^TR^{\left(\mc{R}\right)}E=E^T\LR{E}{R} \:.
%\]
%This means that:
%\[
\begin{equation*}
\mathbf{F}\left(\mc{S}\right)=\tr{E^{T}E-E^TR^{\pth{\sR}}R^{\pth{\sR}}E}=\tr{E^{T}E-\LR{E}{R}\LR{E}{R}}=\|E\|_{F}^{2}-\|\LR{E}{R}\|_{F}^{2}\:.
\end{equation*}
Replacing $\|E\|_{F}^{2}$ with $\mathbf{F}\left(\mc{P}\right)$ proves
the theorem. \hfill\BlackBox

The term $\|\LR{E}{R}\|_{F}^{2}$ represents the
decrease in reconstruction error achieved by adding the subset
$\mc{R}$ of columns to $\mc{P}$. In the following section, a novel
greedy heuristic is presented to optimize the column subset selection
criterion based on this recursive formula.

\subsection{Greedy Selection Algorithm}\label{Sec:CSS_Greedy}

This section presents an efficient greedy algorithm to optimize the
column subset selection criterion presented in Section
\ref{Sec:CSS}. The algorithm selects at each iteration one
column such that the reconstruction error for the new set of
columns is minimized. This problem can be formulated as follows:

\begin{problem}
    At iteration $t$, find column $p$ such that,
    \begin{equation}
        p=\underset{i}{\arg\min}\quad \mathbf{F}\left(\mc{S}\cup\left\{i\right\} \right)
    \end{equation}where $\mc{S}$ is the set of columns selected
    during the first $t-1$ iterations.
\end{problem}

A na\"{i}ve implementation of the greedy algorithm is to calculate
the reconstruction error for each candidate column, and then select
the column with the smallest error. This implementation is, however,
computationally very complex, as it requires $\bigOO(m^2n^2)$
floating-point operations per iteration. A more efficient approach
is to use the recursive formula for calculating the reconstruction
error. Using Theorem \ref{Th:RecF},
\begin{displaymath}
\mathbf{F}\left(\mc{S}\cup\left\{ i\right\}
\right)=\mathbf{F}\left(\mc{S}\right)-\|\tl{E}_{\left\{ i\right\}
}\|_{F}^{2},\end{displaymath}where $E=A-\LR{A}{S}$ and $\tl{E}_{\left\{ i\right\}}$ is the rank-$1$ approximation of $E$ based on the candidate column $i$. Since
$\mathbf{F}\left(\mc{S}\right)$ is a constant for all candidate
columns, an equivalent criterion is:
\begin{equation}\label{Eq:OptGreedyEq}
p=\underset{i}{\arg\max}\hspace{1em}\|\tl{E}_{\left\{ i\right\}
}\|_{F}^{2}
\end{equation}
This formulation selects the column $p$ which achieves the maximum
decrease in reconstruction error. Using the properties that: $\tr{AB}=\tr{BA}$ and $\tr{aA}=a\:\tr{A}$ where $a$ is a scalar, the new objective function
$\left\Vert \tl{E}_{\left\{ i\right\} }\right\Vert _{F}^{2}$ can
be simplified as follows:
\begin{displaymath}
    \begin{split}
        \left\Vert \tl{E}_{\left\{ i\right\} }\right\Vert
        _{F}^{2}&=trace\left(\tl{E}_{\left\{ i\right\}
        }^{T}\tl{E}_{\left\{ i\right\} }\right)=
        trace\left(E^{T}R^{\left(\left\{ i\right\} \right)}E\right)\\
        &=trace\left(E^{T}E_{:i}\left(E_{:i}^{T}E_{:i}\right)^{-1}E_{:i}^{T}E\right)\\
        &=\frac{1}{E_{:i}^{T}E_{:i}}trace\left(E^{T}E_{:i}E_{:i}^{T}E\right)=\frac{\left\Vert
        E^{T}E_{:i}\right\Vert ^{2}}{E_{:i}^{T}E_{:i}}.
    \end{split}
\end{displaymath}

This defines the following equivalent problem.

\begin{problem} {\bf (Greedy Column Subset Selection)}
    At iteration $t$, find column $p$ such that,
    \begin{equation}
        p=\underset{i}{\arg\max}\hspace{1em}\frac{\left\Vert E^{T}E_{:i}\right\Vert ^{2}}{E_{:i}^{T}E_{:i}}
    \end{equation}where $E=A-\LR{A}{S}$, and $\mc{S}$ is the set of columns selected
    during the first $t-1$ iterations.
\end{problem}

The computational complexity of this selection criterion is
$\bigOO\left(n^2m\right)$ per iteration, and it requires
$\bigOO\left(nm\right)$ memory to store the residual of the
whole matrix, $E$, after each iteration.
%In the rest of this section, two novel techniques are proposed to reduce the memory and time requirements of this selection criterion.
% \subsection{Memory-Efficient Criterion} \label{Sec:MemoryCSS}
In order to reduce these memory requirements, a memory-efficient algorithm can be proposed calculate the column subset selection criterion without explicitly calculating and storing the residual matrix $E$ at each iteration. The algorithm is based on a recursive formula for calculating the residual matrix $E$.

Let $\mc{S}^{(t)}$ denote the set of columns selected during the
first $t-1$ iterations, $E^{(t)}$ denote the residual matrix at the
start of the $t$-th iteration (i.e., $E^{(t)}=A -
\tl{A}_{\mc{S}^{(t)}}$), and $p^{(t)}$ be the column selected at
iteration $t$. The following lemma gives a recursive formula for
residual matrix at the start of iteration $t+1$, $E^{(t+1)}$.

\begin{lemma} \label{L_ERec}
    $E^{(t+1)}$ can be calculated
    recursively as:
        \begin{equation} \label{Eq:ERec}
        E^{(t+1)} = \pth{E-\frac{E_{:p}E_{:p}^T}{E_{:p}^TE_{:p}}E}^{(t)}.
        \end{equation}
%          \begin{displaymath}
%        E^{(t+1)} = A - \sum_{r=1}^{t} (\frac{E_{:p}E_{:p}^T}{E_{:p}^TE_{:p}}E)^{(r)}.
%        \end{displaymath}
\end{lemma}
\proof Using Corollary \ref{Cl:LowRank}, $\tl{A}_{\mc{S}
\cup \{p\}}=\LR{A}{S}+\tl{E}_{\{p\}}$. Subtracting
both sides from $A$, and substituting $A-\tl{A}_{\mc{S} \cup
\{p\}}$ and $A-\LR{A}{S}$ with $E^{(t+1)}$ and
$E^{(t)}$ respectively gives:
   \begin{displaymath}
        E^{(t+1)}=\pth{E-\tl{E}_{\{p\}}}^{(t)}
   \end{displaymath}
Using Equations (\ref{Eq:LowRank}) and (\ref{Eq:Projection}),
$\tl{E}_{\{p\}}$ can be expressed as
$\left(E_{:p}(E_{:p}^TE_{:p})^{-1}E_{:p}^T\right)E$. Substituting
$\tl{E}_{\{p\}}$ with this formula in the above equation proves
the lemma. \hfill\BlackBox

%Using Lemma \ref{L_ERec} simplifies the calculation of the residual
%matrix at each iteration. However, this recursive formula still
%needs $\bigOO(mn)$ memory at each iteration.

Let $G$ be an $n \times n$ matrix which represents the inner-products over the columns of the residual matrix $E$: $G = E^TE$. The following
corollary is a direct result of Lemma \ref{L_ERec}.

\begin{corollary} \label{C_GRec}
    $G^{(t+1)}$ can be calculated recursively as:
        \begin{displaymath}
        G^{(t+1)} = \pth{G-\frac{G_{:p}G_{:p}^T}{G_{pp}}}^{(t)}.
        \end{displaymath}
%       \begin{displaymath}
%       G^{(t+1)} = A^TA - \sum_{r=1}^{t} (\frac{G_{:p}G_{:p}^T}{G_{pp}})^{(r)}.
%       \end{displaymath}
\end{corollary}

\proof This corollary can be proved by substituting with
$E^{(t+1)^T}$ (Lemma \ref{L_ERec}) in
$G^{(t+1)}=E^{(t+1)^T}E^{(t+1)}$, and using the fact that $R^{\pth{\{p\}}}R^{\pth{\{p\}}}=R^{\pth{\{p\}}}$ (an idempotent matrix).
%\begin{equation*}
%  \pth{E_{:p}(E_{:p}^TE_{:p})^{-1}E_{:p}^T}\left(E_{:p}(E_{:p}^TE_{:p})^{-1}E_{:p}^T\right)=E_{:p}(E_{:p}^TE_{:p})^{-1}E_{:p}^T\:.
%\end{equation*}
\hfill\BlackBox

To simplify the derivation of the memory-efficient algorithm, at
iteration $t$, define $\bs{\delta} = G_{:p}$ and $\bs{\omega} =
G_{:p}/\sqrt{G_{pp}} = \bs{\delta}/\sqrt{\bs{\delta}_{p}}$. This
means that $G^{(t+1)}$ can be calculated in terms of $G^{(t)}$ and
$\bs{\omega}^{(t)}$ as follows:
\begin{equation} \label{eq:G_RecW}
         G^{(t+1)} = \pth{G-\bs{\omega}\bs{\omega}^{T}}^{(t)},
\end{equation} or in terms of $A$ and previous $\bs{\omega}$'s as:
\begin{equation} \label{eq:G_RecW2}
        G^{(t+1)} = A^TA-\sum_{r=1}^{t} \pth{\bs{\omega}\bs{\omega}^{T}}^{(r)}.
\end{equation}

$\bs{\delta}^{(t)}$ and $ \bs{\omega}^{(t)}$ can be calculated in
terms of $A$ and previous $\bs{\omega}$'s as follows:
 \begin{equation}
     \begin{split} \label{eq:delta_omega}
        \bs{\delta}^{(t)}&=A^{T}A_{:p}-\sum_{r=1}^{t-1}\bs{\omega}_{p}^{(r)}\bs{\omega}^{(r)}, \\
        \bs{\omega}^{(t)}&=\bs{\delta}^{(t)}/\sqrt{\bs{\delta}_{p}^{(t)}}.
    \end{split}
 \end{equation}
%         \begin{displaymath}
%        G^{(t+1)} = A^TA-\sum_{r=1}^{t} (\bs{\omega}\bs{\omega}^{T})^{(r)}.
%        \end{displaymath}

The column subset selection criterion can be expressed in terms of $G$ as:
   \begin{displaymath}
        p=\underset{i}{\arg\max}\hspace{1em}\frac{\left\Vert
        G_{:i}\right\Vert ^{2}}{G_{ii}}
   \end{displaymath}

The following theorem gives recursive formulas for calculating the
column subset selection criterion without explicitly calculating $E$ or
$G$.
\begin{theorem} \label{Th:Rec_fg}
Let $\bs{f}_{i}=\left\Vert  G_{:i} \right\Vert ^{2}$ and
$\bs{g}_{i}=G_{ii}$ be the numerator and denominator of the
criterion function for column $i$ respectively,
$\bs{f}=\left[\bs{f}_{i}\right]_{i=1..n}$, and
$\bs{g}=\left[\bs{g}_{i}\right]_{i=1..n}$. Then,
%\begin{displaymath}
%    \begin{split}
%        \bs{f}^{\left[t\right]}=&\Bigg(\bs{f}-2\left(\omega\circ\left(A^{T}A\omega-\sum_{r=1}^{t-2}\left(\omega^{\left[r\right]T}\omega\right)\omega^{\left[r\right]}\right)\right)\\
%        &+\left\Vert \omega\right\Vert
%        ^{2}\left(\omega\circ\omega\right)\Bigg)^{\left[t-1\right]},\\
%        % \bs{g}^{(t)}&=\Big(\bs{g}-\left(\bs{\omega}\circ\bs{\omega}\right)\Big)^{(t-1)}.
%        \bs{g}^{\left[t\right]}=&\Big(\bs{g}-\omega\circ\omega\Big)^{\left[t-1\right]}
%    \end{split}
%\end{displaymath}
\begin{displaymath}
\begin{split}
\bs{f}^{(t)}&=\Big(\bs{f}-2\left(\bs{\omega}\circ\left(A^TA\bs{\omega}-\Sigma_{r=1}^{t-2}\left(\bs{\omega}^{{\left(r\right)}^T}\bs{\omega}\right)\bs{\omega}^{^{\left(r\right)}}\right)\right)
+\|\bs{\omega}\|^{2}\left(\bs{\omega}\circ\bs{\omega}\right)\Big)^{(t-1)},\\
\bs{g}^{(t)}&=\Big(\bs{g}-\left(\bs{\omega}\circ\bs{\omega}\right)\Big)^{(t-1)}.
\end{split}
\end{displaymath}where $\circ$ represents the Hadamard product
operator.
%, and $\omega$ is an $n \times 1$ vector which can be
%calculated recursively as follows:
%\begin{displaymath}
%\begin{split}
%\omega^{\left[t-1\right]}&=v^{\left[t-1\right]}/\sqrt{v_{p}^{\left[t-1\right]}}
%\\
%v^{\left[t-1\right]}&=A^{T}A_{:p}-\sum_{r=1}^{t-2}\omega_{p}^{\left[r\right]}\omega^{\left[r\right]}
%\end{split}
%\end{displaymath}
\end{theorem}

\proof Based on Equation (\ref{eq:G_RecW}), $\bs{f}_{i}^{(t)}$ can be
calculated as:
\begin{equation}
\begin{split}
\bs{f}_{i}^{\left(t\right)}&=\left(\left\Vert  G_{:i} \right\Vert ^{2}\right)^{(t)}=\left(\|G_{:i}-\bs{\omega}_i\bs{\omega}\|^{2}\right)^{\left(t-1\right)}\\
&=\left((G_{:i}-\bs{\omega}_i\bs{\omega})^T(G_{:i}-\bs{\omega}_i\bs{\omega})\right)^{\left(t-1\right)}\\
&=\left(G_{:i}^TG_{:i}-2\bs{\omega}_{i}G_{:i}^{T}\bs{\omega}+\bs{\omega}_{i}^{2}\|\bs{\omega}\|^{2}\right)^{\left(t-1\right)}\\
&=\left(\bs{f}_{i}-2\bs{\omega}_{i}G_{:i}^{T}\bs{\omega}+\bs{\omega}_{i}^{2}\|\bs{\omega}\|^{2}\right)^{\left(t-1\right)}.
\end{split}
\end{equation}
Similarly, $\bs{g}_{i}^{(t)}$ can be calculated as:
\begin{equation}
\begin{split}
\bs{g}_{i}^{(t)}&=G_{ii}^{(t)}=\left(G_{ii}-\bs{\omega}_{i}^{2}\right)^{\left(t-1\right)}\\
&=\left(\bs{g}_{i}-\bs{\omega}_{i}^{2}\right)^{\left(t-1\right)}.
\end{split}
\end{equation}

Let $\bs{f}=\left[\bs{f}_{i}\right]_{i=1..n}$and
$\bs{g}=\left[\bs{g}_{i}\right]_{i=1..n}$, $\bs{f}^{(t)}$ and
$\bs{g}^{(t)}$ can be expressed as:
\begin{equation}
\begin{split}
\bs{f}^{(t)}&=\left(\bs{f}-2\left(\bs{\omega}\circ
G\bs{\omega}\right)+\|\bs{\omega}\|^{2}\left(\bs{\omega}\circ\bs{\omega}\right)\right)^{(t-1)},
\\
\bs{g}^{(t)}&=\left(\bs{g}-\left(\bs{\omega}\circ\bs{\omega}\right)\right)^{(t-1)}\label{eq:f-1},
\end{split}
\end{equation}
where $\circ$ represents the Hadamard product operator, and $\|.\|$
is the $\ell_2$ norm.

Based on the recursive formula of $G$ (Eq. \ref{eq:G_RecW2}), the
term $G\bs{\omega}$ at iteration $(t-1)$ can be expressed as:
\begin{equation}
\begin{split}
G\bs{\omega}&=\left(A^TA-\Sigma_{r=1}^{t-2}\left(\bs{\omega}\bs{\omega}^{T}\right)^{\left(r\right)}\right)\bs{\omega}\\
&=A^TA\bs{\omega}-\Sigma_{r=1}^{t-2}\left(\bs{\omega}^{{\left(r\right)}^T}\bs{\omega}\right)\bs{\omega}^{^{\left(r\right)}}
\end{split}
\end{equation}

Substituting with $G\bs{\omega}$ in Equation (\ref{eq:f-1}) gives the
update formulas for $\bs{f}$ and $\bs{g}$. \hfill\BlackBox

This means that the greedy criterion can be memory-efficient by only
maintaining two score variables for each column, $\bs{f}_{i}$ and
$\bs{g}_{i}$, and updating them at each iteration based on their
previous values and the columns selected so far.

\begin{algorithm}[t]
\caption{\label{alg:Greedy-Algorithm}Greedy Column Subset Selection}
\textbf{Input:} Data matrix $A$, Number of columns $l$\\
\textbf{Output:} Selected subset of columns $\mc{S}$

%\begin{enumerate}
\begin{algorithmic}[1]
\STATE Initialize
$\mc{S}=\{~\}$
\STATE Initialize $\bs{f}_i^{(0)}=\|A^TA_{:i}\|^{2}$,
$\bs{g}_i^{(0)}=A_{:i}^TA_{:i}$ for $i=1\:...\:n$
\STATE Repeat $t=1\rightarrow l$:
%\begin{enumerate}
\STATE \hspace{0.25cm} $p=\underset{i}{\arg\max}\ \bs{f}_{i}^{(t)}/\bs{g}_{i}^{(t)}$,~~~$\mc{S}=\mc{S}\cup  \{p\}$
\STATE \hspace{0.25cm} $\bs{\delta}^{(t)}=A^TA_{:p}-\sum_{r=1}^{t-1}\bs{\omega}_{p}^{(r)}\bs{\omega}^{(r)}$
\STATE \hspace{0.25cm} $\bs{\omega}^{(t)}=\bs{\delta}^{(t)}/\sqrt{\bs{\delta}^{(t)}_{p}}$
\STATE \hspace{0.25cm} Update $\bs{f}_i$'s, $\bs{g}_i$'s (Theorem \ref{Th:Rec_fg})
%\end{enumerate}
%\item     $W=\left[\begin{array}{cccc}\bs{\omega}^{(1)}&...&\bs{\omega}^{(l)}\end{array}\right]^T$
%\end{enumerate}
\end{algorithmic}
\end{algorithm}

Algorithm \ref{alg:Greedy-Algorithm} shows the complete greedy CSS algorithm. %The distributed CSS algorithm presented in this paper introduces a generalized variant of the greedy CSS algorithm in which a subset of columns is selected from a source matrix such that the reconstruction error of a target matrix is minimized. The distributed CSS method uses the greedy generalized CSS algorithm as the core method for selecting columns at different machines as well as in the final selection step.

\section{Distributed Column Subset Selection on MapReduce} \label{Sec:DistCSS}

This section describes a MapReduce algorithm for the distributed column subset selection problem. Given a big data matrix $A$ whose columns are distributed across different machines, the goal is to select a subset of columns $\mc{S}$ from $A$ such that the CSS criterion $\mathbf{F}\left(\mc{S}\right)$ is minimized.

%As the data matrix $A$ cannot be loaded into the memory of a single machine, the centralized algorithms for CSS such as the greedy algorithm need to be modified to select columns from distributed matrices

%In order to solve this optimization problem using centralized algorithms such as the greedy CSS algorithm, the whole data matrix has to be loaded into the memory of a single machine, which is infeasible for big data matrices. Therefore, a MapReduce algorithm needs to be developed in which different subsets of columns are first selected from different sub-matrices of $A$, and then aggregated at a reducer.

One na\"{\i}ve approach to perform distributed column subset selection is to select different subsets of columns from the sub-matrices stored on different machines. The selected subsets are then sent to a centralized machine where an additional selection step is optionally performed to filter out irrelevant or redundant columns. Let $A_{(i)}$ be the sub-matrix stored at machine $i$, the na\"{\i}ve approach optimizes the following function.
\begin{equation}
\sum_{i=1}^{c}\left\Vert A_{\left(i\right)}-P^{\left(\mc{L}_{\left(i\right)}\right)}A_{\left(i\right)}\right\Vert _{F}^{2}\:,
\end{equation}where $\mc{L}_{(i)}$ is the set of columns selected from $A_{(i)}$ and $c$ is the number of physical blocks of data. The resulting set of columns is the union of the sets selected from different sub-matrices: $\mc{L}=\cup_{i=1}^{c}\mc{L}_{\left(i\right)}$. The set $\mc{L}$ can further be reduced by invoking another selection process in which a smaller subset of columns is selected from $A_{:\mc{L}}$.
%In the na\"{\i}ve approach, the number of columns to be selected from each sub-matrix $l_{\left(i\right)}=\left|\mc{L}_{\left(i\right)}\right|$ has to be provided to the algorithm or it can be set to $l/c$, where $l=\left|\mc{L}\right|$, or more adaptively to $l\left\Vert A_{\left(i\right)}\right\Vert _{F}^{2}/\left\Vert A\right\Vert _{F}^{2}$.

The  na\"{\i}ve  approach, however simple, is prone to missing relevant columns. This is because the selection at each machine is based on approximating a local sub-matrix, and accordingly there is no way to determine whether the selected columns are globally relevant or not. For instance, suppose the extreme case where all the truly representative columns happen to be loaded on a single machine. In this case, the algorithm will select a less-than-required number of columns from that machine and many irrelevant columns from other machines.

In order to alleviate this problem, the different machines have to select columns that best approximate a common representation of the data matrix. To achieve that, the proposed algorithm first learns a concise representation of the span of the big data matrix. This concise representation is relatively small and it can be sent over to all machines. After that each machine can select columns from its sub-matrix that approximate this concise representation. The proposed algorithm uses random projection to learn this concise representation, and proposes a generalized Column Subset Selection (CSS) method to select columns from different machines. The details of the proposed methods are explained in the rest of this section.
%
%
%The development of the aforementioned algorithm requires two questions to be answered: First, what is the concise representation that can be effectively computed over MapReduce? Second, how to select columns from each sub-matrix such that a different common matrix is approximated? The two questions are answered in the following subsections.

\subsection{Random Projection}\label{Sec:RP}
The first step of the proposed algorithm is to learn a concise representation $B$ for a distributed data matrix $A$. In the proposed approach, a random projection method is employed. Random projection \cite{RSA:RSA10073} \cite{rs} \cite{srs} is a well-known technique for dealing with the curse-of-the-dimensionality problem. Let $\Omega$ be a random projection matrix of size $n\times r$, and given a data matrix $X$ of size $m\times n$, the random projection can be calculated as $Y=X\Omega$. It has been shown that applying random projection $\Omega$ to $X$ preserves the pairwise distances between vectors in the row space of $X$ with a high probability \cite{RSA:RSA10073}:
\begin{equation}
\begin{split}
\left(1-\epsilon\right)\left\Vert X_{i:}-X_{j:}\right\Vert &\le\left\Vert X_{i:}\Omega-X_{j:}\Omega\right\Vert \\ &\le\left(1+\epsilon\right)\left\Vert X_{i:}-X_{j:}\right\Vert\:,
\end{split}
\end{equation} where $\epsilon$ is an arbitrarily small factor. % that depends on the structure of the random matrix as well as its size.

Since the CSS criterion $\mathbf{F}\left(\mathcal{S}\right)$ measures the reconstruction error between the big data matrix $A$ and its low-rank approximation $P^{\left(\mathcal{S}\right)}A$, it essentially measures the sum of the distances between the original rows and their approximations. This means that when applying random projection to both $A$ and $P^{\left(\mathcal{S}\right)}A$, %that with the selection of a proper random projection matrix,
the reconstruction error of the original data matrix $A$ will be approximately equal to that of $A\Omega$ when both are approximated using the subset of selected columns:
\begin{equation}
  \|A-\proj{S}A\|_{F}^{2} \approx \|A\Omega-\proj{S}A\Omega\|_{F}^{2} \:.
\end{equation}
So, instead of optimizing $\|A-\proj{S}A\|_{F}^{2}$, the distributed CSS can approximately optimize $\|A\Omega-\proj{S}A\Omega\|_{F}^{2}$.

Let $B = A\Omega$, the distributed column subset selection problem can be formally defined as

\begin{problem} {\bf (Distributed Column Subset Selection)} \label{Pr:GenCSS}
    Given an $m\times n_{(i)}$ sub-matrix $A_{(i)}$ which is stored at node $i$ and an integer $l_{(i)}$, find a subset of columns $\mc{L}_{(i)}$ such that $|\mc{L}_{(i)}| =l_{(i)}$ and
    \begin{displaymath}
        \mc{L}_{(i)}=\underset{\mc{S}}{\arg\min} \|B-\proj{S}B\|_{F}^{2},
    \end{displaymath} where $B = A\Omega$, $\Omega$ is an $n\times r$ random projection matrix, $\mc{S}$ is the set of the indices of the candidate columns and $\mc{L}_{(i)}$ is the set of the indices of the selected columns from $A_{(i)}$.
\end{problem}

A key observation here is that random projection matrices whose entries are sampled $i.i.d$ from some univariate distribution $\Psi$ can be exploited to compute random projection on MapReduce in a very efficient manner. %avoiding the time consuming matrix multiplication on MapReduce.
Examples of such matrices are Gaussian random matrices \cite{RSA:RSA10073}, uniform random sign ($\pm 1$) matrices \cite{rs}, and sparse random sign matrices \cite{srs}. %($\pm 1$ with each with probability $1/2\sqrt{n}$ and $0$ with probability $1- 1/\sqrt{n}$)

In order to implement random projection on MapReduce, the data matrix $A$ is distributed in a column-wise fashion and viewed as pairs of $\langle i, A_{:i}\rangle$ where $A_{:i}$ is the $i$-th column of $A$. Recall that $B = A\Omega$ can be rewritten as
\begin{equation}
B = \sum_{i=1}^{n} A_{:i} \Omega_{i:}
\end{equation} and since the \textit{map} function is provided one columns of $A$ at a time, one does not need to worry about pre-computing the full matrix $\Omega$. In fact, for each input column $A_{:i}$, a new vector $\Omega_{i:}$ needs to be sampled from $\Psi$. So, each input column generates a matrix of size $m \times r$ which means that $O(nmr)$ data should be moved across the network to sum the generated $n$ matrices at $m$ independent $reducers$ each summing a row $B_{j:}$ to obtain $B$. To minimize that network cost, an \textit{in-memory} summation can be carried out over the generated $m \times r$ matrices at each mapper. This can be done incrementally after processing each column of $A$. That optimization reduces the network cost to $O(cmr)$, where $c$ is the number of physical blocks of the matrix\footnote{The \textit{in-memory} summation can also be replaced by a MapReduce \textit{combiner} \cite{Dean2008}.}. Algorithm \ref{Alg:RndProj} outlines the proposed random projection algorithm. %The notations \textit{map} and \textit{reduce} are used to refer to the \textit{map} and \textit{reduce} functions of MapReduce.
The term $emit$ is used to refer to outputting new $\langle key,value\rangle$ pairs from a mapper or a reducer.

\begin{algorithm}[t]\caption{Fast Random Projection on MapReduce}\label{Alg:RndProj}
\textbf{Input:} Data matrix $A$, Univariate distribution $\Psi$, Number of dimensions $r$ \\
\textbf{Output:} Concise representation $B=A\Omega,\:\Omega_{ij}\sim\Psi\:\forall i,j$

\begin{algorithmic}[1]
	\STATE \textbf{\textit{map:}}
	\STATE \vspace{0.1cm}\hspace{0.25cm} $\bar{B} = [0]_{m \times r}$
	\STATE \hspace{0.25cm} \textbf{foreach} $\langle i, A_{:i}\rangle$
	\STATE \hspace{0.75cm} Generate $\textbf{v} = [\textbf{v}_1, \textbf{v}_2, ... \textbf{v}_r]$, $\textbf{v}_j\sim\Psi$
	\STATE \hspace{0.75cm} $\bar{B} = \bar{B} + A_{:i}\textbf{v}$	
	\STATE \hspace{0.25cm} \textbf{for} $j$ = $1$ \textbf{to} $m$
	\STATE  \hspace{0.75cm} emit $\langle j, \bar{B}_{j:}\rangle$
	\STATE \vspace{0.2cm}\textbf{\textit{reduce:}}	
	\STATE \vspace{0.1cm}\hspace{0.25cm} \textbf{foreach} $\langle j, \left[[\bar{B}_{(1)}]_{j:},[\bar{B}_{(2)}]_{j:},...,[\bar{B}_{(c)}]_{j:}\right]\rangle$
	\STATE \hspace{0.75cm} $B_{j:} = \sum_{i=1}^{c} [\bar{B}_{(i)}]_{j:}$
	\STATE \vspace{0.1cm}\hspace{0.75cm} emit $\langle j, B_{j:} \rangle$
\end{algorithmic}
%\textbf{Input:} Matrix $A$ of size $m \times n$, Univariate distribution $\Psi$, Number of dimensions $r$ \\
%\textbf{Output:} Subspace embedding $B= A\Omega$ where $\Omega_{ij} \sim \Psi \: \forall i,j$
%\begin{algorithmic}[1]
%	\STATE $\hat{B} \leftarrow 0_{r \times d}$
%	\STATE \textbf{map}$\langle i, A_{:i}\rangle$
%	\STATE \hspace{0.25cm} sample $v_{1 \times d}$ such that $v(i) \sim \Psi \forall i \in 1,2,..r$
%	\STATE \hspace{0.25cm} $\hat{B} \leftarrow \hat{B} + A_{:i} \times v$	
%	\STATE \textbf{finmap}
%	\STATE \hspace{0.25cm} \textbf{for} $i$ = 1 \textbf{to} \($d$\)
%	\STATE  \hspace{0.5cm} emit $\langle i, \hat{B}_{:i}\rangle$
%	\STATE \textbf{reduce} $\langle i, b_1, b_2,..b_P\rangle$
%	\STATE  \hspace{0.25cm} $B_{:i} \leftarrow \sum_{j=1}^{P} b_j$
%	\STATE  \vspace{0.1cm}\hspace{0.25cm} emit $\langle i, B_{:i}\rangle$
%\end{algorithmic}
\end{algorithm}

\subsection{Generalized Column Subset Selection}
This section presents the generalized column subset selection algorithm which will be used to perform the selection of columns at different machines. While Problem \ref{Pr:FS} is concerned with the selection of a subset of columns from a data matrix which best represent other columns of the same matrix, Problem \ref{Pr:GenCSS} selects a subset of columns from a source matrix which best represent the columns of a different target matrix. The objective function of Problem \ref{Pr:GenCSS} represents the reconstruction error of the target matrix $B$ based on the selected columns from the source matrix. and the term $P^{\left(\mc{S}\right)}=A_{:\mc{S}} \inv{A_{:\mc{S}}^{T}A_{:\mc{S}}} A_{:\mc{S}}^{T}$ is the projection matrix which projects the columns of $B$ onto the subspace of the columns selected from $A$.

In order to optimize this new criterion, a greedy algorithm can be introduced. Let $\mathbf{\bar{F}}\left(\mathcal{S}\right)=\left\Vert B-P^{\left(\mathcal{S}\right)}B\right\Vert _{F}^{2}$ be the distributed CSS criterion, the following theorem derives a recursive formula for $\mathbf{\bar{F}}\left(\mathcal{S}\right)$.

\begin{theorem}\label{Th:RecF}
    Given a set of columns $\mc{S}$. For any $\mc{P} \subset \mc{S}$,
    \begin{displaymath}
        \mathbf{\bar{F}}\left(\mathcal{S}\right)=\mathbf{\bar{F}}\left(\mathcal{P}\right)-\left\Vert \tilde{F}_{\mathcal{R}}\right\Vert _{F}^{2}
 \:,
    \end{displaymath} where $F = B - P^{\left(\mc{P}\right)}B$, and $\LR{F}{R}$ is the low-rank
    approximation of $F$ based on the subset $\mc{R} = \mc{S}\setminus \mc{P}$ of
    columns of $E=A - P^{\left(\mc{P}\right)}A$.
\end{theorem}
\proof Using the recursive formula for the low-rank approximation of $A$: $\tilde{A}_{\mathcal{S}}=\tilde{A}_{\mathcal{P}}+\tilde{E}_{\mathcal{R}}$, and multiplying both sides with $\Omega$ gives
\begin{equation*}
  \tilde{A}_{\mathcal{S}}\Omega=\tilde{A}_{\mathcal{P}}\Omega+\tilde{E}_{\mathcal{R}}\Omega\:.
\end{equation*}
Low-rank approximations can be written in terms of projection matrices as
\begin{equation*}
  P^{\left(\mathcal{S}\right)}A\Omega=P^{\left(\mathcal{P}\right)}A\Omega+R^{\left(\mathcal{R}\right)}E\Omega\:.
\end{equation*}
Using $B=A\Omega$,
\begin{equation*}
   P^{\left(\mathcal{S}\right)}B=P^{\left(\mathcal{P}\right)}B+R^{\left(\mathcal{R}\right)}E\Omega\:.
\end{equation*}
Let $F=E\Omega$. The matrix $F$ is the residual after approximating $B$ using the set $\mathcal{P}$ of columns
\begin{equation*}
  F=E\Omega=\left(A-P^{\left(\mathcal{P}\right)}A\right)\Omega=A\Omega-P^{\left(\mathcal{P}\right)}A\Omega=B-P^{\left(\mathcal{P}\right)}B\:.
\end{equation*}
This means that
\begin{equation*}
  P^{\left(\mathcal{S}\right)}B=P^{\left(\mathcal{P}\right)}B+R^{\left(\mathcal{R}\right)}F
\end{equation*}
Substituting in $\mathbf{\bar{F}}\left(\mathcal{S}\right)=\left\Vert B-P^{\left(\mathcal{S}\right)}B\right\Vert _{F}^{2}$ gives
\begin{equation*}
  \mathbf{\bar{F}}\left(\mathcal{S}\right)=\left\Vert B-P^{\left(\mathcal{P}\right)}B-R^{\left(\mathcal{R}\right)}F\right\Vert _{F}^{2}
\end{equation*}
Using $F=B-P^{\left(\mathcal{P}\right)}B$ gives
\begin{equation*}
    \mathbf{\bar{F}}\left(\mathcal{S}\right)=\left\Vert F-R^{\left(\mathcal{R}\right)}F\right\Vert _{F}^{2}
\end{equation*}
Using the relation between Frobenius norm and trace,
\begin{displaymath}
\begin{split}
  \mathbf{\bar{F}}\left(\mathcal{S}\right)&=\text{trace}\left(\left(F-R^{\left(\mathcal{R}\right)}F\right)^{T}\left(F-R^{\left(\mathcal{R}\right)}F\right)\right)\\
  &=\text{trace}\left(F^{T}F-2F^{T}R^{\left(\mathcal{R}\right)}F+F^{T}R^{\left(\mathcal{R}\right)}R^{\left(\mathcal{R}\right)}F\right) \\
  &=\text{trace}\left(F^{T}F-F^{T}R^{\left(\mathcal{R}\right)}F\right)=\left\Vert F\right\Vert _{F}^{2}-\left\Vert R^{\left(\mathcal{R}\right)}F\right\Vert _{F}^{2}
\end{split}
\end{displaymath}
Using $\mathbf{\bar{F}}\left(\mathcal{P}\right)=\left\Vert F\right\Vert _{F}^{2}$ and $\tilde{F}_{\mathcal{R}}=R^{\left(\mathcal{R}\right)}F$ proves the theorem. \hfill\BlackBox

\begin{algorithm}[t]
\caption{\label{Alg:GenGCSS}Greedy Generalized
Column Subset Selection}
\textbf{Input:} Source matrix $A$, Target matrix $B$, Number of columns $l$\\
\textbf{Output:} Selected subset of columns $\mc{S}$

\begin{algorithmic}[1]
%\begin{enumerate}
\STATE Initialize $\bs{f}_i^{(0)}=\|B^TA_{:i}\|^{2}$,
$\bs{g}_i^{(0)}=A_{:i}^TA_{:i}$ for $i=1\:...\:n$
\STATE Repeat $t=1\rightarrow l$:
%\begin{enumerate}
\STATE \hspace{0.25cm} $p=\underset{i}{\arg\max}\ \bs{f}_{i}^{(t)}/\bs{g}_{i}^{(t)}$,~~~$\mc{S}=\mc{S}\cup  \{p\}$
\STATE \hspace{0.25cm}
$\bs{\delta}^{(t)}=A^TA_{:p}-\sum_{r=1}^{t-1}\bs{\omega}_{p}^{(r)}\bs{\omega}^{(r)}$
\STATE \hspace{0.25cm} $\bs{\gamma}^{(t)}=B^TA_{:p}-\sum_{r=1}^{t-1}\bs{\omega}_{p}^{(r)}\bs{\upsilon}^{(r)}$
\STATE \hspace{0.25cm} $\bs{\omega}^{(t)}=\bs{\delta}^{(t)}/\sqrt{\bs{\delta}^{(t)}_{p}}$, $\bs{\upsilon}^{(t)}=\bs{\gamma}^{(t)}/\sqrt{\bs{\delta}^{(t)}_{p}}$
\STATE \hspace{0.25cm} Update $\bs{f}_i$'s, $\bs{g}_i$'s (Theorem \ref{Th:Rec_fg_2})
%\begin{displaymath}
%\begin{split}\bs{f}^{(t)}&=\Big(\bs{f}-2\left(\bs{\omega}\circ\left(A^{T}B\bs{\upsilon}-\Sigma_{r=1}^{t-2}\left(\bs{\upsilon}^{\left(r\right)T}\bs{\upsilon}\right)\bs{\omega}^{^{\left(r\right)}}\right)\right)
%+\|\bs{\upsilon}\|^{2}\left(\bs{\omega}\circ\bs{\omega}\right)\Big)^{(t-1)},\\
%\bs{g}^{(t)}
%&=\Big(\bs{g}-\left(\bs{\omega}\circ\bs{\omega}\right)\Big)^{(t-1)}\:,\end{split}
%\end{displaymath} where $\circ$ represents the Hadamard product operator.
%\end{enumerate}
%\end{enumerate}
\end{algorithmic}
\end{algorithm}

Using the recursive formula for $\mathbf{\bar{F}}\left(\mathcal{S}\cup\{i\}\right)$ allows the development of a greedy algorithm which at iteration $t$ optimizes
\begin{equation}
  p=\underset{i}{\arg\min}\:\mathbf{\bar{F}}\left(\mathcal{S}\cup\{i\}\right)=\underset{i}{\arg\max}\left\Vert \tilde{F}_{\left\{ i\right\} }\right\Vert _{F}^{2}
\end{equation}

Let $G=E^TE$ and $H=F^TE$, the objective function of this optimization problem can be simplified as follows.
\begin{equation}
\begin{split}
\left\Vert \tilde{F}_{\left\{ i\right\} }\right\Vert _{F}^{2}&=\left\Vert E_{:i}\left(E_{:i}^{T}E_{:i}\right)^{-1}E_{:i}^{T}F\right\Vert _{F}^{2}\\
&=\text{trace}\left(F^TE_{:i}\left(E_{:i}^{T}E_{:i}\right)^{-1}E_{:i}^{T}F\right)\\
&=\frac{\left\Vert F^TE_{:i}\right\Vert ^{2}}{E_{:i}^{T}E_{:i}}=\frac{\left\Vert H_{:i}\right\Vert ^{2}}{G_{ii}}\:.
\end{split}
\end{equation}

This allows the definition of the following generalized CSS problem.
\begin{problem} {\bf (Greedy Generalized CSS)} \label{Pr:GreedyCSS}
    At iteration $t$, find column $p$ such that
    \begin{equation*}
        p=\underset{i}{\arg\max}\hspace{1em}\frac{\left\Vert H_{:i}\right\Vert ^{2}}{G_{ii}}
    \end{equation*}where $H=F^TE$, $G = E^TE$, $F=B-P^{\left(\mathcal{S}\right)}B$, $E=A-P^{\left(\mathcal{S}\right)}A$ and $\mc{S}$ is the set of columns selected during the first $t-1$ iterations.
\end{problem}

For iteration $t$, define $\bs{\gamma} = H_{:p}$ and $\bs{\upsilon} =
H_{:p}/\sqrt{G_{pp}} = \bs{\gamma}/\sqrt{\bs{\delta}_{p}}$ \:. The vector $\bs{\gamma}^{(t)}$ can be calculated in terms of $A$, $B$ and previous $\bs{\omega}$'s and $\bs{\upsilon}$'s
as
%\begin{equation} \label{eq:delta_omega}
 $\bs{\gamma}^{(t)}=B^{T}A_{:p}-\sum_{r=1}^{t-1}\bs{\omega}_{p}^{(r)}\bs{\upsilon}^{(r)}\:$.
%\end{equation}

%Similarly, the numerator and denominator of the selection criterion at each iteration can be calculated in an efficient manner without explicitly calculating $F$ or $H$ using the following theorem.
Similarly, the numerator and denominator of the selection criterion at each iteration can be calculated in an efficient manner using the following theorem.
\begin{theorem} \label{Th:Rec_fg_2}
Let $\bs{f}_{i}=\left\Vert H_{:i}\right\Vert ^{2}$ and
$\bs{g}_{i}=G_{ii}$ be the numerator and denominator of the
greedy criterion function for column $i$ respectively,
$\bs{f}=\left[\bs{f}_{i}\right]_{i=1..n}$, and
$\bs{g}=\left[\bs{g}_{i}\right]_{i=1..n}$. Then,
\begin{displaymath}
\begin{split}\bs{f}^{(t)}&=\Big(\bs{f}-2\left(\bs{\omega}\circ\left(A^{T}B\bs{\upsilon}-\Sigma_{r=1}^{t-2}\left(\bs{\upsilon}^{\left(r\right)T}\bs{\upsilon}\right)\bs{\omega}^{^{\left(r\right)}}\right)\right)+\|\bs{\upsilon}\|^{2}\left(\bs{\omega}\circ\bs{\omega}\right)\Big)^{(t-1)},\\
\bs{g}^{(t)}
&=\Big(\bs{g}-\left(\bs{\omega}\circ\bs{\omega}\right)\Big)^{(t-1)}\:,\end{split}
\end{displaymath} where $\circ$ represents the Hadamard product operator.
\end{theorem}

\begin{algorithm} [t] \caption{\label{Alg:GenGCSSMap}Distributed CSS on MapReduce}
\textbf{Input:} Matrix $A$ of size $m \times n$, Concise representation $B$, Number of columns $l$\\
\textbf{Output:} Selected columns $C$

\begin{algorithmic}[1]
	\STATE \textbf{\textit{map:}}
    \STATE \vspace{0.1cm}\hspace{0.25cm}$A_{(b)} = [$ $]$
    \STATE \hspace{0.25cm}\textbf{foreach} $\langle i, A_{:i}\rangle$
    \STATE \hspace{0.75cm} $A_{(b)} = [A_{(b)}$ $A_{:i}]$
    \STATE \hspace{0.25cm} $ \bar{\mathcal{S}} =  $ GeneralizedCSS($A_{(b)},B,l_{(b)}$)
    \STATE \hspace{0.25cm} \textbf{foreach} $j$ \textbf{in} $\bar{\mathcal{S}}$
    \STATE \hspace{0.75cm} emit $\langle 0, [A_{(b)}]_{:j} \rangle$
    \STATE \vspace{0.2cm}\textbf{\textit{reduce:}}	
	\STATE \vspace{0.1cm}\hspace{0.25cm} For all values $ \{ [A_{(1)}]_{:\bar{\mathcal{S}}_{(1)}}, [A_{(2)}]_{:\bar{\mathcal{S}}_{(2)}}, ...., [A_{(c)}]_{:\bar{\mathcal{S}}_{(c)}}\}$
% $[\bs{a}^{(1)},\bs{a}^{(2)},...\bs{a}^{(\sum_{b=1}^{c} l_{(b)})}]$
    \STATE \hspace{0.25cm} $A_{(0)}=$ $ \left[ [A_{(1)}]_{:\bar{\mathcal{S}}_{(1)}}, [A_{(2)}]_{:\bar{\mathcal{S}}_{(2)}}, ...., [A_{(c)}]_{:\bar{\mathcal{S}}_{(c)}}\right]$
    %    $[\bs{a}^{(1)},\bs{a}^{(2)},...\bs{a}^{(\sum_{b=1}^{c} l_{(b)})}]$
    \STATE \hspace{0.25cm} $\mathcal{S} = $ GeneralizedCSS ($A_{(0)}$, $B$, $l$)
    \STATE \hspace{0.25cm} \textbf{foreach} $j$ \textbf{in} $\mathcal{S}$
    \STATE \hspace{0.75cm} emit $\langle0, [A_{(0)}]_{:j}\rangle$
\end{algorithmic}
\end{algorithm}

As outlined in Section \ref{Sec:RP}, the algorithm's distribution strategy is based on sharing the concise representation of the data $B$ among all mappers. Then, independent $l_{(b)}$ columns from each mapper are selected using the generalized CSS algorithm. A second phase of selection is run over the $\sum_{b=1}^{c} l_{(b)}$ (where $c$ is the number of input blocks) columns to find the best $l$ columns to represent $B$. Different ways can be used to set $l_{(b)}$ for each input block $b$. In the context of this paper, the set of $l_{(b)}$ is assigned uniform values for all blocks (i.e. $l_{(b)} = \lfloor l/c \rfloor \forall b \in 1,2, .. c$). Algorithm \ref{Alg:GenGCSSMap} sketches the MapReduce implementation of the distributed CSS algorithm. It should be emphasized that the proposed MapReduce algorithm requires only two passes over the data set and its moves a very few amount of the data across the network.

\section{Related Work} \label{Sec:RelatedWork}

Different approaches have been proposed for selecting a subset of representative columns from a data matrix. This section focuses on briefly describing these approaches and their applicability to massively distributed data matrices. The Column Subset Selection (CSS) methods can be generally categorized into randomized, deterministic and hybrid.

\subsection{Randomized Methods}

The randomized methods sample a subset of columns from the original matrix using carefully chosen sampling probabilities. The main focus of this category of methods is to develop fast algorithms for column subset selection and then derive a bound for the reconstruction error of the data matrix based on the selected columns relative to the best possible reconstruction error obtained using Singular Value Decomposition (SVD).

Frieze et al. \cite{Frieze98-Rnd} was the first to suggest the idea of randomly sampling $l$ columns from a matrix and using these columns to calculate a rank-$k$ approximation of the matrix (where $l\ge k$). The authors derived an additive bound for the reconstruction error of the data matrix. This work of Frieze et al. was followed by different papers \cite{Drineas04-clust} \cite{Drineas2007a} that enhanced the algorithm by proposing different sampling probabilities and deriving better error bounds for the reconstruction error. Drineas et al. \cite{Drineas06-Cols} proposed a subspace sampling method which samples columns using probabilities proportional to the norms of the rows of the top $k$ right singular vectors of $A$. The subspace sampling method allows the development of a relative-error bound (i.e., a multiplicative error bound relative to the best rank-$k$ approximation). However, the subspace sampling depends on calculating the leading singular vectors of a matrix which is computationally very complex for large  matrices.

Deshpande et al. \cite{Deshpande06a-Vol} \cite{Deshpande06b-Vol} proposed an adaptive sampling method which updates the sampling probabilities based on the columns selected so far. This method is computationally very complex, as it depends on calculating the residual of the data matrix after each iteration. In the same paper, Deshpande et al. also proved the existence of a volume sampling algorithm (i.e., sampling a subset of columns based on the volume enclosed by their vectors) which achieves a multiplicative ($l+1$)-approximation. However, the authors did not present a polynomial time algorithm for this volume sampling algorithm.

%The randomized methods sample a subset of columns from the original matrix using carefully chosen sampling probabilities. Frieze et al. \cite{Frieze98-Rnd} was the first to suggest the idea of randomly sampling $l$ columns from a matrix and using these columns to calculate a rank-$k$ approximation of the matrix (where $l\ge k$). That work of Frieze et al. was followed by different papers \cite{Drineas04-clust} \cite{Drineas2007a} that enhanced the algorithm by proposing different sampling probabilities. Drineas et al. \cite{Drineas06-Cols} proposed a subspace sampling method which samples columns using probabilities proportional to the norms of the rows of the top $k$ right singular vectors of $A$.  Deshpande et al. \cite{Deshpande06b-Vol} proposed an adaptive sampling method which updates the sampling probabilities based on the columns selected so far.

Column subset selection with uniform sampling can be easily implemented on MapReduce. For non-uniform sampling, the efficiency of implementing the selection on MapReduce is determined by how easy are the calculations of the sampling probabilities. The calculations of probabilities that depend on calculating the leading singular values and vectors are time-consuming on MapReduce. On the other hand, adaptive sampling methods are computationally very complex as they depend on calculating the residual of the whole data matrix after each iteration.

\subsection{Deterministic Methods}

The second category of methods employs a deterministic algorithm for selecting columns such that some criterion function is minimized. This criterion function usually quantifies the reconstruction error of the data matrix based on the subset of selected columns. The deterministic methods are slower, but more accurate, than the randomized ones.

In the area of numerical linear algebra, the column pivoting method exploited by the QR decomposition \cite{Golub1996} permutes the columns of the matrix based on their norms to enhance the numerical stability of the QR decomposition algorithm. The first $l$ columns of the permuted matrix can be directly selected as representative columns. The Rank-Revealing QR (RRQR) decomposition \cite{chan1987rank} \cite{Gu96-QR} \cite{bischof1998computing} \cite{pan2000existence} is a category of QR decomposition methods which permute columns of the data matrix while imposing additional constraints on the singular values of the two sub-matrices of the upper-triangular matrix $R$ corresponding to the selected and non-selected columns. It has been shown that the constrains on the singular values can be used to derive an theoretical guarantee for the column-based reconstruction error according to spectral norm \cite{Boutsidis08-Clust}.

Besides methods based on QR decomposition, different recent methods have been proposed for directly selecting a subset of columns from the data matrix. Boutsidis et al. \cite{Boutsidis08-Clust} proposed a deterministic column subset selection method which first groups columns into clusters and then selects a subset of columns from each cluster. The authors proposed a general framework in which different clustering and subset selection algorithms can be employed to select a subset of representative columns. {\c{C}}ivril and Magdon-Ismail \cite{civril2008deter} \cite{Civril12-CSS-Sparse} presented a deterministic algorithm which greedily selects columns from the data matrix that best represent the right leading singular values of the matrix. This algorithm, however accurate, depends on the calculation of the leading singular vectors of a matrix, which is computationally very complex for large matrices.

Recently, Boutsidis et al. \cite{Boutsidis11a-NearOpt} presented a column subset selection algorithm which first calculates the top-$k$ right singular values of the data matrix (where $k$ is the target rank) and then uses deterministic sparsification methods to select $l\ge k$ columns from the data matrix. The authors derived a theoretically near-optimal error bound for the rank-$k$ column-based approximation. Deshpande and Rademacher \cite{Deshpande10-Vol} presented a polynomial-time deterministic algorithm for volume sampling with a theoretical guarantee for $l=k$. Quite recently, Guruswami and Sinop \cite{Guruswami12-Optimal} presented a deterministic algorithm for volume sampling with theoretical guarantee for $l>k$. The deterministic volume sampling algorithms are, however, more complex than the algorithms presented in this paper, and they are infeasible for large data sets.

The deterministic algorithms are more complex to implement on MapReduce. For instance, it is time-consuming to calculate the leading singular values and vectors of a massively distributed matrix or to cluster their columns using $k$-means. It is also computationally complex to calculate QR decomposition with pivoting. Moreover, the recently proposed algorithms for volume sampling are more complex than other CSS algorithms as well as the one presented in this paper, and they are infeasible for large data sets.

%Some recent work proposes greedy algorithm for subset selection. This includes the work of \cite{DasK11} \cite{CevherK11}

\subsection{Hybrid Methods}

A third category of CSS techniques is the hybrid methods which combine the benefits of both the randomized and deterministic methods. In these methods, a large subset of columns is randomly sampled from the columns of the data matrix and then a deterministic step is employed to reduce the number of selected columns to the desired rank.

For instance, Boutsidis et al. \cite{Boutsidis09a-CSS} proposed a two-stage hybrid algorithm for column subset selection which runs in $\bigO{\min\pth{n^2m, nm^2}}$. In the first stage, the algorithm samples $c=\bigO{l\log l}$ columns based on probabilities calculated using the $l$-leading right singular vectors. In the second phase, a Rank-revealing QR (RRQR) algorithm is employed to select exactly $l$ columns from the columns sampled in the first stage. The authors suggested repeating the selection process 40 times in order to provably reduce the failure probability. The authors proved a good theoretical guarantee for the algorithm in terms of spectral and Frobenius term. However, the algorithm depends on calculating the leading $l$ right singular vectors which is computationally complex for large data sets.

The hybrid algorithms for CSS can be easily implemented on MapReduce if the randomized selection step is MapReduce-efficient and the deterministic selection step can be implemented on a single machine. This is usually true if the number of columns selected by the randomized step is relatively small.

\subsection{Comparison to Related Work}
The greedy column subset selection algorithm presented in Section \ref{Sec:GreedyCSS} belongs to the category of deterministic algorithms. In comparison to QR-based methods, the greedy CSS algorithm can be implicitly used to calculate a $Q$-less incomplete $QR$ factorization of the data matrix $A$:
\begin{equation*}
A=QW, \:  A\Pi=QW\Pi=QR
\end{equation*} where $\Pi$ is a permutation matrix which sorts the first $l$ columns according to their selection order. The permutation of the columns of the embedding matrix $W$ produces an upper triangular matrix.

The greedy CSS algorithm differs from the greedy algorithm proposed by  {\c{C}}ivril and Magdon-Ismail \cite{civril2008deter} \cite{Civril12-CSS-Sparse} in that the latter depends on first calculating the Singular Value Decomposition of the data matrix, which is computationally complex, especially for large matrices. The proposed algorithm is also more efficient than the recently proposed volume sampling algorithms \cite{Deshpande10-Vol} \cite{Guruswami12-Optimal}.

%On the other hand, the distributed greedy CSS algorithm presented in Section \ref{Sec:DistCSS} belongs to the category of hybrid algorithms in the sense that it combines a randomized step with a deterministic selection. However, the randomized step in the distributed algorithm depends on random projection rather than selecting a set random samples; this random projection speeds up the algorithm without losing much information about the span of the columns.

In comparison to other CSS methods, the distributed algorithm proposed in this paper is designed to be MapReduce-efficient. In the selection step, representative columns are selected based on a common representation. The common representation proposed in this work is based on random projection. This is more efficient than the work of {\c{C}}ivril and Magdon-Ismail \cite{Civril12-CSS-Sparse} which selects columns based on the leading singular vectors. In comparison to other deterministic methods, the proposed algorithm is specifically designed to be parallelized which makes it applicable to big data matrices whose columns are massively distributed. On the other hand, the two-step of distributed then centralized selection is similar to that of the hybrid CSS methods. The proposed algorithm however employs a deterministic algorithm at the distributed selection phase which is more accurate than the randomized selection employed by hybrid methods in the first phase.

\begin{table}
\begin{center}
\caption{\label{tab:Data sets-1}The properties of the medium an large data sets used to evaluate different CSS methods.}
\begin{tabular}{|c||c|c|c|}
\hline \textbf{Data set} & \textbf{Type} & \textbf{\# Instances} & \textbf{\# Features} \tabularnewline
\hline \hline Reuters-21578 & Documents & 5,946 &
18,933\tabularnewline \hline Reviews & Documents & 4,069 &
36,746\tabularnewline \hline LA1 & Documents & 3,204 &
29,714\tabularnewline \hline MNIST-4K & Digit Images & 4,000 &
784\tabularnewline \hline PIE-20 & Face Images & 3,400 &
1,024\tabularnewline \hline YaleB-38 & Face Images & 2,414 &
1,024\tabularnewline \hline
RCV1-200K & Documents  & 193,844 & 47,236
\tabularnewline \hline
TinyImages-1M & Images  & 1 million & 1,024
\tabularnewline \hline
\end{tabular}
\end{center}
\end{table}

%\section{Experiments and Results}\label{Sec:CSS_Exp}

\section{Experiments} \label{Sec:Experiments}
Two sets of experiments have been conducted. The first set of experiments has been conducted on medium-sized data sets in order to evaluate the efficiency and effectiveness of the  centralized greedy CSS algorithm in comparison to state-of-the-art methods for CSS. The second set of experiments has been conducted on two big data sets to evaluate the efficiency and effectiveness of the distributed CSS algorithm on MapReduce.

Experiments have been conducted on eight
benchmark data sets, whose properties are summarized in Table
\ref{tab:Data sets-1}.\footnote{
The data sets \textit{Reuters-21578}, \textit{MNIST-4K},
\textit{PIE-20} and \textit{YaleB-38} are available in MAT format
at: \url{http://www.cad.zju.edu.cn/home/dengcai/Data/data.html}.
\textit{PIE-20} is a subset of \textit{PIE-32x32} with the images of
the first 20 persons.}
The first six data sets were used to conduct the centralized experiments.
The \textit{Reuters-21578} is the training set of the Reuters-21578
collection \cite{lewis1999rtc}. The \textit{Reviews} and
\textit{LA1} are document data sets from TREC
collections.\footnote{\label{fn:TREC.-Text-REtrieval}\url{http://trec.nist.gov}}
The pre-processed versions of \textit{Reviews} and \textit{LA1} that
are distributed with the CLUTO Toolkit \cite{karypis2002cct} were
used. The \textit{MNIST-4K} is a subset of the MNIST data set of
handwritten digits.\footnote{\url{http://yann.lecun.com/exdb/mnist}}
The \textit{PIE-20} and \textit{YaleB-38} are pre-processed subsets
of the CMU PIE \cite{sim2003cmu} and Extended Yale Face
\cite{KCLee05} data sets respectively. The  \textit{PIE-20} and \textit{YaleB-38} data sets have been used by He et al. \cite{he2005face} to evaluate different face recognition algorithms. Besides, the distributed experiments were conducted on two data sets. The \textit{RCV1-200K} is a subset of the RCV1 data set \cite{lewis2004rcv1} which has been prepared and used by Chen et al. \cite{chen2010parallel} to evaluate parallel spectral clustering algorithms. The \textit{TinyImages-1M} data set contains 1 million images that were sampled from the 80 million tiny images data set \cite{TinyImages} and converted to grayscale.

Similar to previous work on CSS, the different methods are evaluated according to their ability to minimize the reconstruction error of the data matrix based on the subset of selected columns. In order to quantify the reconstruction error across different data sets, a relative accuracy measure is defined as
%a measure that calculates the approximation accuracy relative to uniform sampling is defined as
\begin{equation*}
\text{Relative Accuracy}=
\frac{\Vert A-\LR{A}{U}\Vert_F - \Vert A-\LR{A}{S} \Vert_F}
{\Vert A-\LR{A}{U}\Vert_F - \Vert A-\tl{A}_{l}\Vert_F} \times 100\%\: ,
\end{equation*}where $\LR{A}{U}$ is the rank-$l$ approximation of the data matrix based on a random subset $\mc{U}$ of columns, $\LR{A}{S}$ is the rank-$l$ approximation of the data matrix based on the subset $\sS$ of columns and $\tl{A}_{l}$ is the best rank-$l$ approximation of the data matrix calculated using the Singular Value Decomposition (SVD). This measure compares different methods relative to the uniform sampling as a baseline with higher values indicating better performance.

\subsection{Evaluation of Centralized Greedy CSS}
In the medium-scale experiments, the following CSS methods are compared\footnote{The CSS algorithm of Boutsidis et al. \cite{Boutsidis11a-NearOpt} was not included in the comparison as its implementation is not available.}.
\begin{itemize}
  \item \textbf{UniNoRep}: is uniform sampling of columns without replacement.
  \item \textbf{qr}: is the QR decomposition with column pivoting \cite{Golub1996} implemented by the MATLAB $qr$ function.\footnote{Revision: 5.13.4.7}
  \item \textbf{SRRQR}: is the strong rank-revealing QR decomposition \cite{Gu96-QR}. Algorithm 4 of \cite{Gu96-QR} was implemented in MATLAB. In this implementation, the MATLAB $qr$ function is first used to calculate the QR decomposition with column pivoting and then the columns are swapped using the criterion specified by Gu and Eisenstat \cite{Gu96-QR}.\footnote{In the implemented code, the efficient recursive formulas in Section 4 of \cite{Gu96-QR} are used to implement the update of QR decomposition and the swapping criterion.}
  \item \textbf{ApproxSVD}: is the sparse approximation of Singular Value Decomposition (SVD) \cite{civril2008deter} \cite{Civril12-CSS-Sparse}. The algorithm was implemented in MATLAB. The generalized CSS algorithm is used to select columns that best approximates the leading singular vectors. The use of the generalized CSS algorithm is equivalent to, but more efficient than, the algorithm proposed by {\c{C}}ivril and Magdon-Ismail \cite{civril2008deter} \cite{Civril12-CSS-Sparse}. Since the calculation of exact SVD is computationally complex, the Stochastic SVD algorithm \cite{Halko11a-Survey} is used to approximate the leading singular values and vectors of the data matrix. This significantly reduces the run time of the original algorithm proposed by {\c{C}}ivril and Magdon-Ismail while achieving comparable accuracy. In this experiment, the number of leading singular vectors is set to $l$.
  \item \textbf{HybridCSS}: is the hybrid column subset selection algorithm proposed by Bousidis et al. \cite{Boutsidis09a-CSS}. The number of selected columns in the randomized phase is set to $l\:\log{\pth{l}}$. The algorithm was implemented in MATLAB. In the randomized phase, the Stochastic SVD is first used to calculate the leading singular vectors, and the approximated singular vectors are then used to calculate the sampling probabilities. In the random phase, the number of leading singular vectors is set to $l$. In the deterministic phase, the MATLAB $qr$ function is used to select columns.\footnote{In \cite{Boutsidis09a-CSS-New} (a newer version of \cite{Boutsidis09a-CSS}), Boutsidis et al. suggested the use of the SRRQR algorithm \cite[Algorithm 4]{Gu96-QR} for the deterministic phase. Although the SRRQR algorithm achieves the theoretical guarantee presented in \cite{Boutsidis09a-CSS}, the MATLAB $qr$ function is used in the conducted experiments as it is much faster and it achieves comparable accuracy for the experimented data sets.}
  \item \textbf{GreedyCSS}: is the greedy column subset selection method described in Algorithm \ref{alg:Greedy-Algorithm}.
  \item \textbf{RndGreedyCSS}: is the greedy algorithm for the generalized column subset selection in which the target matrix is a random subspace obtained using random projection. Similar to ApproxSVD and HybridCSS, the dimension of the random projection matrix is set to $l$.
\end{itemize}

\begin{figure*}
\centering
\includegraphics[width=0.9\linewidth]{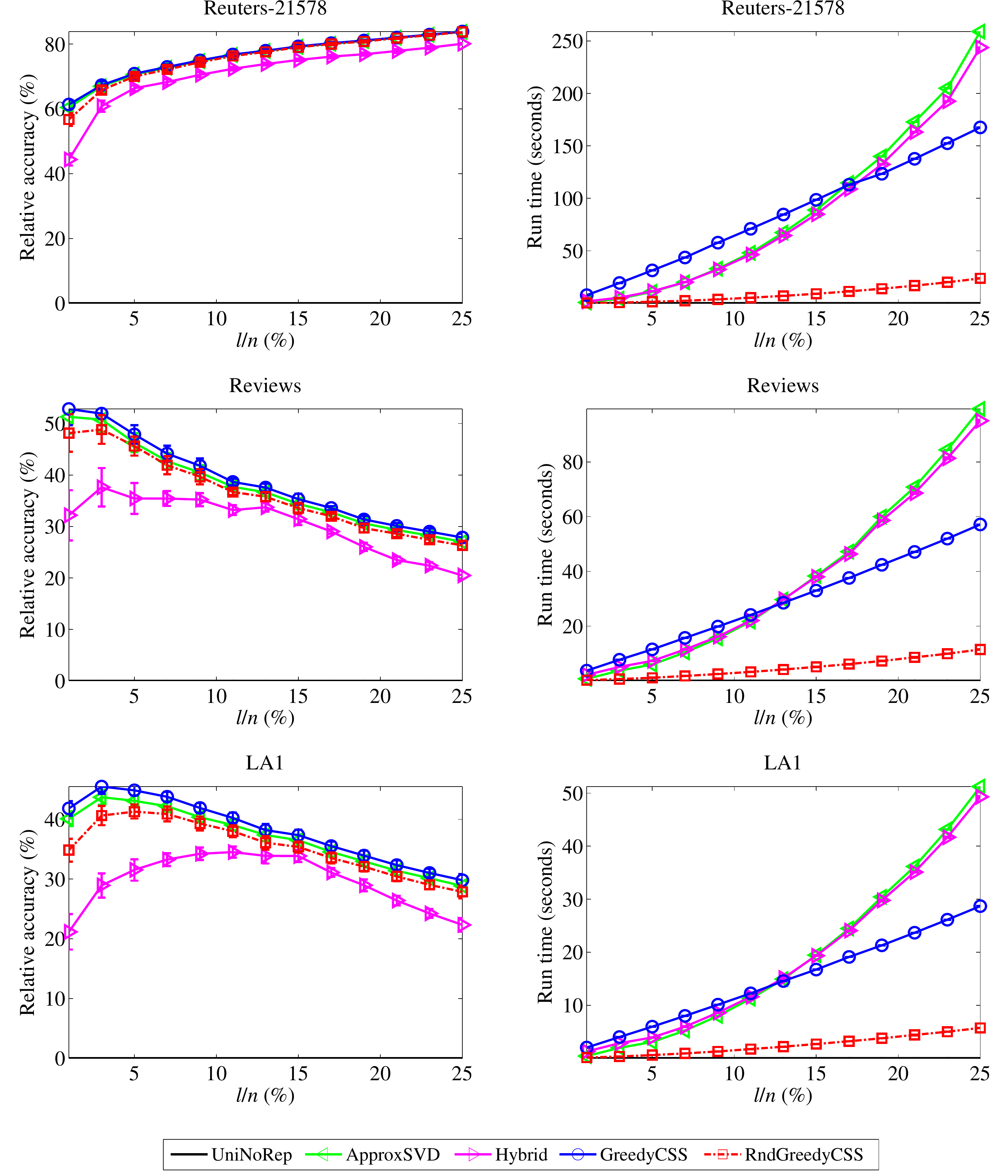}
\caption{\label{fig:Results_CSS1}The relative accuracy measures and run times of different column-based low-rank approximations $\tl{A}_{\mc{S}}$ for the \textit{Reuters-21578}, \textit{Reviews} and
\textit{LA1} data sets.}
\end{figure*}

\begin{figure*}
\centering
\includegraphics[width=0.9\linewidth]{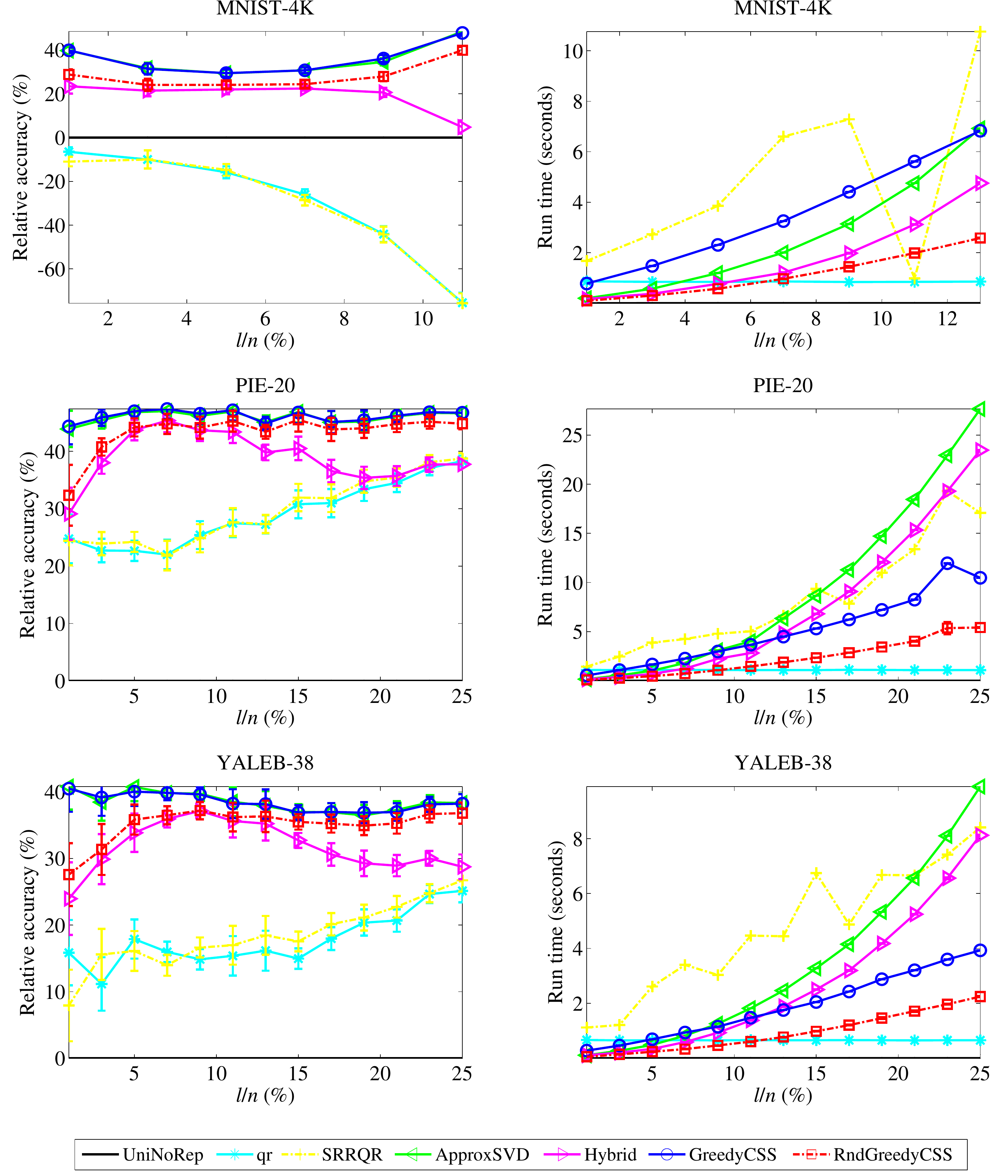}
\caption{\label{fig:Results_CSS2}The relative accuracy measures and run times of different column-based low-rank approximations $\tl{A}_{\mc{S}}$ for the \textit{MNIST-4K},
\textit{PIE-20} and \textit{YaleB-38} data sets.}
\end{figure*}
%The different CSS methods are evaluated according to their ability to minimize the reconstruction error of the data matrix based on the subset of selected columns (Definition \ref{Def:Criterion}). In order to quantify the reconstruction error across different data sets, a relative accuracy measure is defined as
%\begin{equation}
%\text{Relative Accuracy }= \frac{\Vert A-\tl{A}_{l}\Vert_F}{\Vert
%A-\LR{A}{S} \Vert_F} \: ,
%\end{equation}where $\LR{A}{S}$ is the rank-$l$ approximation of the data matrix calculated based on the subset $\sS$ of columns and $\tl{A}_{l}$ is the best rank-$l$ approximation of the data matrix calculated using the Singular Value Decomposition (SVD). Since $\Vert A-\tl{A}_{l}\Vert_F$ is the minimum possible reconstruction error for a rank-$l$ approximation, the relative accuracy is between 0 and 1 with higher values indicating better CSS methods.

For all the data sets, the percentage of selected columns $l/n$ is changed from $1\%$ to $25\%$ with increments of $2\%$ and the relative accuracies and run times are measured.\footnote{For the MNIST4K data set, the range of $l/n$ values is smaller since the rank of the matrix is very low (i.e., less than the number of pixels).} Experiments with randomness were repeated ten times, and the average and standard deviation of measures were calculated.

Figures \ref{fig:Results_CSS1} and \ref{fig:Results_CSS2} show the relative accuracy measures and run times for different CSS methods on the six benchmark data sets.\footnote{The \textbf{qr} and \textbf{SRRQR} methods both depend on the MATLAB qr function. For the document data sets, the MATLAB qr function takes very long times compared to other methods and accordingly they are not reported in the shown figures.}

It can be observed from the figures and tables that for all data sets, the \textbf{GreedyCSS} method significantly outperforms the \textbf{UniNoRep}, \textbf{qr}, \textbf{SRRQR}, and \textbf{HybridCSS} methods in terms of relative accuracy, and it shows comparable accuracy to the \textbf{ApproxSVD} method. In terms of run times, for most of the data sets, the \textbf{GreedyCSS} scales better than the \textbf{HybridCSS} and \textbf{ApproxSVD} methods.

On the other hand, the  \textbf{RndGreedyCSS} outperforms the \textbf{UniNoRep}, \textbf{qr}, and \textbf{SRRQR} methods in terms of relative accuracy, and shows comparable accuracy to the \textbf{HybridCSS} method. In terms of run times, the \textbf{RndGreedyCSS} is much more efficient than the \textbf{HybridCSS} method and other methods for all data sets. It should also be noted that the \textbf{SRRQR} method achieves comparable accuracy to the \textbf{qr} method and both methods demonstrate lower approximation accuracies than other deterministic and hybrid methods.

\subsection{Evaluation of Distributed Greedy CSS}

The distributed CSS method has been compared with different state-of-the-art methods. It should be noted that most of these methods were not designed with the goal of applying them to massively-distributed data, and hence their implementation on MapReduce is not straightforward. However, the designed experiments used the best practices for implementing the different steps of these methods on MapReduce to the best of the authors' knowledge. In specific, the following distributed CSS algorithms were compared.
%\footnote{The CSS algorithm of Boutsidis et al. \cite{Boutsidis11a-NearOpt} was not included in the comparison as its implementation is not available.}
\begin{itemize}
  \item \textbf{UniNoRep}: is uniform sampling of columns without replacement. This is usually the worst performing method in terms on approximation error and it will be used as a baseline to evaluate methods across different data sets.
%  \item \textbf{qr}: is the QR decomposition with column pivoting \cite{Golub1996} implemented by the MATLAB $qr$ function.\footnote{Revision: 5.13.4.7}
%  \item \textbf{SRRQR}: is the strong rank-revealing QR decomposition \cite{Gu96-QR}. Algorithm 4 of \cite{Gu96-QR} was implemented in MATLAB. In this implementation, the MATLAB $qr$ function is first used to calculate the QR decomposition with column pivoting and then the columns are swapped using the criterion specified by Gu and Eisenstat \cite{Gu96-QR}.\footnote{In the implemented code, the efficient recursive formulas in Section 4 of \cite{Gu96-QR} are used to implement the update of QR decomposition and the swapping criterion.}

  \item \textbf{HybirdUni}, \textbf{HybirdCol} and \textbf{HybirdSVD}: are different distributed variants of the hybrid CSS algorithm which can be implemented efficiently on MapReduce. In the randomized phase, the three methods use probabilities calculated based on uniform sampling, column norms and the norms of the leading singular vectors' rows, respectively. The number of selected columns in the randomized phase is set to $l\:\log{\pth{l}}$. In the deterministic phase, the centralized greedy CSS is employed to select exactly $l$ columns from the randomly sampled columns. %\footnote{The deterministic approach suggested by Bousidis et al. \cite{Boutsidis09a-CSS} is very complex to implement on MapReduce.}
      %In the deterministic phase, the MATLAB $qr$ function is used to select columns.\footnote{In \cite{Boutsidis09a-CSS-New} (a newer version of \cite{Boutsidis09a-CSS}), Boutsidis et al. suggested the use of the SRRQR algorithm \cite[Algorithm 4]{Gu96-QR} for the deterministic phase. Although the SRRQR algorithm achieves the theoretical guarantee presented in \cite{Boutsidis09a-CSS}, the MATLAB $qr$ function is used in the conducted experiments as it is much faster and it achieves comparable accuracy for the experimented data sets.}
%  \item \textbf{GreedyCSS}: is the greedy column subset selection method described in Algorithm \ref{alg:Greedy-Algorithm}.
  \item \textbf{DistApproxSVD}: is an extension of the centralized algorithm for sparse approximation of Singular Value Decomposition (SVD) \cite{Civril12-CSS-Sparse}.  The distributed CSS algorithm presented in this paper (Algorithm \ref{Alg:GenGCSSMap}) is used to select columns that best approximate the leading singular vectors (by setting $B=U_k\Sigma_k$). The use of the distributed CSS algorithm extends the original algorithm proposed by {\c{C}}ivril and Magdon-Ismail \cite{Civril12-CSS-Sparse} to work on distributed matrices. In order to allow efficient implementation on MapReduce, the number of leading singular vectors is set of $100$.
  \item \textbf{DistGreedyCSS}: is the distributed column subset selection method described in Algorithm \ref{Alg:GenGCSSMap}. For all experiments, the dimension of the random projection matrix is set to $100$. This makes the size of the concise representation the same as the DistApproxSVD method. Two types of random matrices are used for random projection: (1) a dense Gaussian random matrix (rnd), and (2) a sparse random sign matrix (ssgn).
\end{itemize}

For the methods that require the calculations of Singular Value Decomposition (SVD), the Stochastic SVD (SSVD) algorithm \cite{Halko11a-Survey} is used to approximate the leading singular values and vectors of the data matrix. The use of SSVD significantly reduces the run time of the original SVD-based algorithms while achieving comparable accuracy. In the conducted experiments, the SSVD implementation of Mahout was used.

The distributed experiments were conducted on Amazon EC2\footnote{Amazon Elastic Compute Cloud (EC2): \url{http://aws.amazon.com/ec2}} clusters, which consist of 10 instances for the \textit{RCV1-200K} data set and 20 instances for the \textit{TinyImages-1M} data set. Each instance has a 7.5 GB of memory and a two-cores processor. All instances are running Debian 6.0.5 and Hadoop version 1.0.3. The data sets were converted into a binary format in the form of a sequence of \textit{key-value} pairs. Each pair consisted of a column index as the key and a vector of the column entries. That is the standard format used in Mahout\footnote{Mahout is an Apache project for implementing Machine Learning algorithms on Hadoop. See \url{http://mahout.apache.org/}.} for storing distributed matrices.

Table \ref{tab:Results1} shows the run times and relative accuracies for different CSS methods. It can be observed from the table that for the \textit{RCV1-200K} data set, the DistGreedyCSS methods (with random Gaussian and sparse random sing matrices) outperforms all other methods in terms of relative accuracies. In addition, the run times of both of them are relatively small compared to the DistApproxSVD method which achieves accuracies that are close to the DistGreedyCSS method. Both the DistApproxSVD and DistGreedyCSS methods achieve very good approximation accuracies compared to randomized and hybrid methods. It should also be noted that using a sparse random sign matrix for random projection takes much less time than a dense Gaussian matrix, while achieving comparable approximation accuracies. Based on this observation, the sparse random matrix has been used with the \textit{TinyImages-1M} data set.

For the \textit{TinyImages-1M} data set, although the DistApproxSVD achieves slightly higher approximation accuracies than DistGreedyCSS (with sparse random sign matrix), the DistGreedyCSS selects columns in almost one-third of the time. The reason why the DistApproxSVD outperforms DistGreedyCSS for this data set is that its rank is relatively small (less than 1024). This means that using the leading 100 singular values to represent the concise representation of the data matrix captures most of the information in the matrix and accordingly is more accurate than random projection. The DistGreedyCSS however still selects a very good subset of columns in a relatively small time.

\begin{table}
\begin{center}
\caption{\label{tab:Results1}The run times and relative accuracies of different distributed CSS methods. The best performing method for each $l$ is highlighted in bold, and the second best method is underlined. Negative measures indicate methods that perform worse than uniform sampling.}

\begin{tabular}{|c||c|c|c||c|c|c|}
\hline
\multirow{2}{*}{Methods} & \multicolumn{3}{c||}{\textbf{Run time (minutes)}} & \multicolumn{3}{c|}{\textbf{Relative accuracy (\%)}}\tabularnewline
\cline{2-7}
 & $l=10$ & $l=100$ & $l=500$ & $l=10$ & $l=100$ & $l=500$\tabularnewline
\hline
\multicolumn{7}{|c|}{\textbf{RCV1 - 200K}}\tabularnewline
\hline
\textbf{Uniform - Baseline} & 0.6 & 0.6 & 0.5 & 0.00 & 0.00 & 0.00\tabularnewline
\hline
\textbf{Hybird (}\textbf{\footnotesize Uniform}\textbf{)} & 0.8 & 0.8 & 2.9 & -2.37 & -1.28 & 4.49\tabularnewline
\hline
\textbf{Hybird (}\textbf{\footnotesize Column Norms}\textbf{)} & 1.6 & 1.5 & 3.7 & 4.54 & 0.81 & 6.60\tabularnewline
\hline
\textbf{Hybird (}\textbf{\footnotesize SVD-based}\textbf{)} & 1.3 & 1.4 & 3.6 & 9.00 & 12.10 & 18.43\tabularnewline
\hline
\textbf{Distributed Approx. SVD } & 16.6 & 16.7 & 18.8 & \underline{41.50} & 57.19 & 63.10\tabularnewline
\hline
\textbf{Distributed Greedy CSS (rnd)} & 5.8 & 6.2 & 7.9 & \textbf{51.76} & \underline{61.92} & \underline{67.75}\tabularnewline
\hline
\textbf{Distributed Greedy CSS (ssgn)} & 2.2 & 2.9 & 5.1 & 40.30 & \textbf{62.41} & \textbf{67.91}\tabularnewline
\hline
\multicolumn{7}{|c|}{\textbf{Tiny Images - 1M}}\tabularnewline
\hline
\textbf{Uniform - Baseline} & 1.3 & 1.3 & 1.3 & 0.00 & 0.00 & 0.00\tabularnewline
\hline
\textbf{Hybird (}\textbf{\footnotesize Uniform}\textbf{)} & 1.5 & 1.7 & 8.3 & 19.99 & 6.85 & 6.50\tabularnewline
\hline
\textbf{Hybird (}\textbf{\footnotesize Column Norms}\textbf{)} & 3.3 & 3.4 & 9.4 & 17.28 & 3.57 & 7.80\tabularnewline
\hline
\textbf{Hybird (}\textbf{\footnotesize SVD-based}\textbf{)} & 52.4 & 52.5 & 59.4 & 3.59 & 8.57 & 10.82\tabularnewline
\hline
\textbf{Distributed Approx. SVD } & 71.0 & 70.8 & 75.2 & \textbf{70.02} & \textbf{31.05} & \textbf{24.49}\tabularnewline
\hline
\textbf{Distributed Greedy CSS (ssgn)} & 22.1 & 23.6 & 24.2 & \underline{67.58} & \underline{25.18} & \underline{20.74}\tabularnewline
\hline
\end{tabular}

\end{center}
\end{table}

%It should be noted that the relative accuracies for the images data sets are generally higher than those of the documents data sets. This is due to the fact that the rank of the data matrices of the document data sets is very large compared to the rank of the matrices of the image data sets (which is less than the number of pixels in each image).

% Tables \ref{tab:Results1} and \ref{tab:Results2}

%\begin{tabular}{|c||c|c|c||c|c|c|}
%\hline
%\multirow{2}{*}{Methods} & \multicolumn{3}{c||}{Run time (seconds)} & \multicolumn{3}{c|}{Relative accuracy (\%)}\tabularnewline
%\cline{2-7}
% & $l=10$ & $l=50$ & $l=100$ & $l=10$ & $l=50$ & $l=100$\tabularnewline
%\hline
%\hline
%UniNoRep & 33 & 33 & 38 & 24.48 & 37.15 & 48.84\tabularnewline
%\hline
%HybirdU ({\footnotesize Uniform}) & 50 & 41 & 47 & 20.06 & 37.42 & 45.36\tabularnewline
%\hline
%HybirdN ({\footnotesize Column Norms}) & 93 & 88 & 91 & 26.01 & 33.76 & 47.26\tabularnewline
%\hline
%HybirdL ({\footnotesize Leverage}) & 78 & 93 & 81 & 29.78 & 42.58 & 52.89\tabularnewline
%\hline
%DistributedApproxSVD  & 994 & 993 & 1003 & \textbf{49.80} & 68.88 & 75.53\tabularnewline
%\hline
%DistributedGreedyCSS & 349 & 357 & 370 & 34.96 & \textbf{70.57} & \textbf{77.54}\tabularnewline
%\hline
%\end{tabular}
%
%\end{center}
%\end{table*}
%
%\begin{table*}
%\begin{center}
%\caption{\label{tab:Results2}The run times and relative accuracies of different methods on the ??? data set.}
%\end{center}
%\end{table*}

\section{Conclusion} \label{Sec:Conclusion}

This paper proposes a novel algorithm which greedily selects a subset of columns from a data matrix such that reconstruction error of the data matrix is minimized. The algorithm depends on a novel recursive formula for the reconstruction error of the data matrix, which allows a greedy selection criterion to be calculated efficiently at each iteration. This paper also presents an accurate and efficient MapReduce algorithm for selecting a subset of columns from a massively distributed matrix. The algorithm starts by learning a concise representation of the data matrix using random projection. It then selects columns from each sub-matrix that best approximate this concise approximation. A centralized selection step is then performed on the columns selected from different sub-matrices. In order to facilitate the implementation of the proposed method, a novel algorithm for greedy generalized CSS is proposed to perform the selection from different sub-matrices. In addition, the different steps of the algorithms are carefully designed to be MapReduce-efficient. Experiments on medium and big data sets demonstrate the effectiveness and efficiency of the proposed algorithm in comparison to other CSS methods when implemented on centralized and distributed data.

\vskip 0.2in
% BibTeX users please use one of
%\bibliographystyle{spbasic}      % basic style, author-year citations
%\bibliographystyle{spmpsci}      % mathematics and physical sciences
%\bibliographystyle{spphys}       % APS-like style for physics
%\bibliography{}   % name your BibTeX data base

%\bibliographystyle{IEEEtran}
\bibliographystyle{spmpsci}
\bibliography{References}

\begin{thebibliography}{10}
\providecommand{\url}[1]{{#1}}
\providecommand{\urlprefix}{URL }
\expandafter\ifx\csname urlstyle\endcsname\relax
  \providecommand{\doi}[1]{DOI~\discretionary{}{}{}#1}\else
  \providecommand{\doi}{DOI~\discretionary{}{}{}\begingroup
  \urlstyle{rm}\Url}\fi

\bibitem{rs}
Achlioptas, D.: {Database-friendly random projections: Johnson-Lindenstrauss
  with binary coins}.
\newblock Journal of computer and System Sciences \textbf{66}(4), 671--687
  (2003)

\bibitem{bischof1998computing}
Bischof, C., Quintana-Ort{\'\i}, G.: {Computing rank-revealing QR
  factorizations of dense matrices}.
\newblock ACM Transactions on Mathematical Software (TOMS) \textbf{24}(2),
  226--253 (1998)

\bibitem{Boutsidis11a-NearOpt}
Boutsidis, C., Drineas, P., Magdon-Ismail, M.: Near optimal column-based matrix
  reconstruction.
\newblock In: Proceedings of the 52nd Annual IEEE Symposium on Foundations of
  Computer Science (FOCS'11), pp. 305 --314 (2011).
\newblock \doi{10.1109/FOCS.2011.21}

\bibitem{Boutsidis09a-CSS-New}
Boutsidis, C., Mahoney, M.W., Drineas, P.: An improved approximation algorithm
  for the column subset selection problem.
\newblock CoRR \textbf{abs/0812.4293} (2008)

\bibitem{Boutsidis09a-CSS}
Boutsidis, C., Mahoney, M.W., Drineas, P.: An improved approximation algorithm
  for the column subset selection problem.
\newblock In: Proceedings of the Twentieth Annual {ACM-SIAM} Symposium on
  Discrete Algorithms ({SODA}'09), pp. 968--977 (2009)

\bibitem{Boutsidis08-Clust}
Boutsidis, C., Sun, J., Anerousis, N.: Clustered subset selection and its
  applications on it service metrics.
\newblock In: Proceedings of the Seventeenth {ACM} Conference on Information
  and Knowledge Management ({CIKM}'08), pp. 599--608 (2008).
\newblock \doi{10.1145/1458082.1458162}

\bibitem{civril2008deter}
\c{C}ivril, A., Magdon-Ismail, M.: {Deterministic sparse column based matrix
  reconstruction via greedy approximation of SVD}.
\newblock In: Proceedings of the 19th International Symposium on Algorithms and
  Computation (ISAAC'08), pp. 414--423. Springer-Verlag (2008)

\bibitem{Civril12-CSS-Sparse}
\c{C}ivril, A., Magdon-Ismail, M.: {Column subset selection via sparse
  approximation of SVD}.
\newblock Theoretical Computer Science \textbf{421}(0), 1 -- 14 (2012).
\newblock \doi{10.1016/j.tcs.2011.11.019}

\bibitem{chan1987rank}
Chan, T.: {Rank revealing QR factorizations}.
\newblock Linear Algebra and Its Applications \textbf{88}, 67--82 (1987)

\bibitem{chen2010parallel}
Chen, W.Y., Song, Y., Bai, H., Lin, C.J., Chang, E.: Parallel spectral
  clustering in distributed systems.
\newblock Pattern Analysis and Machine Intelligence, IEEE Transactions on
  \textbf{33}(3), 568 --586 (2011).
\newblock \doi{10.1109/TPAMI.2010.88}

\bibitem{RSA:RSA10073}
Dasgupta, S., Gupta, A.: {An elementary proof of a theorem of Johnson and
  Lindenstrauss}.
\newblock Random Structures and Algorithms \textbf{22}(1), 60--65 (2003)

\bibitem{Dean2008}
Dean, J., Ghemawat, S.: {MapReduce: Simplified data processing on large
  clusters}.
\newblock Communications of the ACM \textbf{51}(1), 107--113 (2008).
\newblock \doi{10.1145/1327452.1327492}

\bibitem{deerwester1990ils}
Deerwester, S., Dumais, S., Furnas, G., Landauer, T., Harshman, R.: {Indexing
  by latent semantic analysis}.
\newblock Journal of the American Society for Information Science and
  Technology \textbf{41}(6), 391--407 (1990)

\bibitem{Deshpande10-Vol}
Deshpande, A., Rademacher, L.: Efficient volume sampling for row/column subset
  selection.
\newblock In: Proceedings of the 51st Annual IEEE Symposium on Foundations of
  Computer Science (FOCS'10), pp. 329 --338 (2010).
\newblock \doi{10.1109/FOCS.2010.38}

\bibitem{Deshpande06b-Vol}
Deshpande, A., Rademacher, L., Vempala, S., Wang, G.: Matrix approximation and
  projective clustering via volume sampling.
\newblock Theory of Computing \textbf{2}(1), 225--247 (2006).
\newblock \doi{10.4086/toc.2006.v002a012}

\bibitem{Deshpande06a-Vol}
Deshpande, A., Rademacher, L., Vempala, S., Wang, G.: Matrix approximation and
  projective clustering via volume sampling.
\newblock In: Proceedings of the Seventeenth Annual ACM-SIAM Symposium on
  Discrete Algorithms (SODA'06), pp. 1117--1126. ACM, New York, NY, USA (2006).
\newblock \doi{10.1145/1109557.1109681}

\bibitem{Drineas04-clust}
Drineas, P., Frieze, A., Kannan, R., Vempala, S., Vinay, V.: Clustering large
  graphs via the singular value decomposition.
\newblock Machine Learning \textbf{56}(1-3), 9--33 (2004)

\bibitem{Drineas2007a}
Drineas, P., Kannan, R., Mahoney, M.: {Fast Monte Carlo algorithms for matrices
  II: Computing a low-rank approximation to a matrix}.
\newblock SIAM Journal on Computing \textbf{36}(1), 158--183 (2007)

\bibitem{Drineas06-Cols}
Drineas, P., Mahoney, M., Muthukrishnan, S.: Subspace sampling and
  relative-error matrix approximation: Column-based methods.
\newblock In: Approximation, Randomization, and Combinatorial Optimization.
  Algorithms and Techniques, pp. 316--326. Springer Berlin / Heidelberg (2006)

\bibitem{elgoharyCoRR13}
Elgohary, A., Farahat, A.K., Kamel, M.S., Karray, F.: Embed and conquer:
  Scalable embeddings for kernel k-means on mapreduce.
\newblock CoRR \textbf{abs/1311.2334} (2013)

\bibitem{elsayed}
Elsayed, T., Lin, J., Oard, D.W.: {Pairwise document similarity in large
  collections with MapReduce}.
\newblock In: Proceedings of the 46th Annual Meeting of the Association for
  Computational Linguistics on Human Language Technologies: Short Papers
  (HLT'08), pp. 265--268 (2008)

\bibitem{fastclustering}
Ene, A., Im, S., Moseley, B.: {Fast clustering using MapReduce}.
\newblock In: Proceedings of the Seventeenth ACM SIGKDD International
  Conference on Knowledge Discovery and Data Mining (KDD'11), pp. 681--689
  (2011)

\bibitem{farahat13-icdm}
Farahat, A.K., Elgohary, A., Ghodsi, A., Kamel, M.S.: Distributed column subset
  selection on mapreduce.
\newblock In: Proceedings of the Thirteenth IEEE International Conference on
  Data Mining (ICDM'13) (2013).
\newblock To appear

\bibitem{farahat11-icdm}
Farahat, A.K., Ghodsi, A., Kamel, M.S.: An efficient greedy method for
  unsupervised feature selection.
\newblock In: Proceedings of the Eleventh IEEE International Conference on Data
  Mining (ICDM'11), pp. 161 --170 (2011).
\newblock \doi{10.1109/ICDM.2011.22}

\bibitem{farahat12}
Farahat, A.K., Ghodsi, A., Kamel, M.S.: Efficient greedy feature selection for
  unsupervised learning.
\newblock Knowledge and Information Systems \textbf{35}(2), 285--310 (2013)

\bibitem{Frieze98-Rnd}
Frieze, A., Kannan, R., Vempala, S.: {Fast Monte-Carlo algorithms for finding
  low-rank approximations}.
\newblock In: Proceedings of the 39th Annual IEEE Symposium on Foundations of
  Computer Science (FOCS'98), pp. 370 --378 (1998).
\newblock \doi{10.1109/SFCS.1998.743487}

\bibitem{Golub1996}
Golub, G., Van~Loan, C.: Matrix Computations, 3rd edn.
\newblock Johns Hopkins Univ Pr (1996)

\bibitem{Gu96-QR}
Gu, M., Eisenstat, S.C.: {Efficient algorithms for computing a strong
  rank-revealing QR factorization}.
\newblock SIAM Journal on Scientific Computing \textbf{17}(4), 848--869 (1996).
\newblock \doi{10.1137/0917055}

\bibitem{Guruswami12-Optimal}
Guruswami, V., Sinop, A.K.: Optimal column-based low-rank matrix
  reconstruction.
\newblock In: Proceedings of the 21st Annual ACM-SIAM Symposium on Discrete
  Algorithms (SODA'12), pp. 1207--1214 (2012)

\bibitem{Halko11a-Survey}
Halko, N., Martinsson, P.G., Shkolnisky, Y., Tygert, M.: An algorithm for the
  principal component analysis of large data sets.
\newblock SIAM Journal on Scientific Computing \textbf{33}(5), 2580--2594
  (2011).
\newblock \doi{10.1137/100804139}

\bibitem{he2005face}
He, X., Yan, S., Hu, Y., Niyogi, P., Zhang, H.: {Face recognition using
  Laplacianfaces}.
\newblock Pattern Analysis and Machine Intelligence, IEEE Transactions on
  \textbf{27}(3), 328--340 (2005)

\bibitem{Jain88}
Jain, A.K., Dubes, R.C.: Algorithms for Clustering Data.
\newblock Prentice-Hall, Inc., Upper Saddle River, NJ, USA (1988)

\bibitem{hadi2008}
Kang, U., Tsourakakis, C., Appel, A., Faloutsos, C., Leskovec, J.: Hadi: Fast
  diameter estimation and mining in massive graphs with hadoop.
\newblock CMU-ML-08-117  (2008)

\bibitem{mrc}
Karloff, H., Suri, S., Vassilvitskii, S.: {A model of computation for
  MapReduce}.
\newblock In: Proceedings of the 21st Annual ACM-SIAM Symposium on Discrete
  Algorithms (SODA'10), pp. 938--948 (2010)

\bibitem{karypis2002cct}
Karypis, G.: {CLUTO} - a clustering toolkit.
\newblock Tech. Rep. \#02-017, University of Minnesota, Department of Computer
  Science (2003)

\bibitem{kaufman1987cmm}
Kaufman, L., Rousseeuw, P.: {Clustering by means of medoids}.
\newblock Tech. rep., Technische Hogeschool, Delft (Netherlands). Department of
  Mathematics and Informatics (1987)

\bibitem{KCLee05}
Lee, K., Ho, J., Kriegman, D.: Acquiring linear subspaces for face recognition
  under variable lighting.
\newblock Pattern Analysis and Machine Intelligence, IEEE Transactions on
  \textbf{27}(5), 684--698 (2005)

\bibitem{lewis1999rtc}
Lewis, D.: {Reuters-21578 text categorization test collection distribution 1.0}
  (1999)

\bibitem{lewis2004rcv1}
Lewis, D.D., Yang, Y., Rose, T.G., Li, F.: Rcv1: A new benchmark collection for
  text categorization research.
\newblock The Journal of Machine Learning Research \textbf{5}, 361--397 (2004)

\bibitem{srs}
Li, P., Hastie, T.J., Church, K.W.: Very sparse random projections.
\newblock In: Proceedings of the Twelfth ACM SIGKDD international conference on
  Knowledge Discovery and Data Mining (KDD'06), pp. 287--296 (2006).
\newblock \doi{10.1145/1150402.1150436}

\bibitem{lütkepohl1996handbook}
L\"utkepohl, H.: {Handbook of Matrices}.
\newblock John Wiley \& Sons Inc (1996)

\bibitem{robustRegression}
Meng, X., Mahoney, M.: Robust regression on mapreduce.
\newblock In: Proceedings of the 30th International Conference on Machine
  Learning (ICML-13), pp. 888--896 (2013)

\bibitem{pan2000existence}
Pan, C.: {On the existence and computation of rank-revealing LU
  factorizations}.
\newblock Linear Algebra and its Applications \textbf{316}(1), 199--222 (2000)

\bibitem{sim2003cmu}
Sim, T., Baker, S., Bsat, M.: {The CMU pose, illumination, and expression
  database}.
\newblock Pattern Analysis and Machine Intelligence, IEEE Transactions on
  \textbf{25}(12), 1615--1618 (2003)

\bibitem{signhLogistc}
Singh, S., Kubica, J., Larsen, S., Sorokina, D.: Parallel large scale feature
  selection for logistic regression.
\newblock Proceedings of the SIAM International Conference on Data Mining pp.
  1171--1182 (2009)

\bibitem{TinyImages}
Torralba, A., Fergus, R., Freeman, W.: 80 million tiny images: A large data set
  for nonparametric object and scene recognition.
\newblock Pattern Analysis and Machine Intelligence, IEEE Transactions on
  \textbf{30}(11), 1958--1970 (2008)

\bibitem{hadoop}
White, T.: {Hadoop: The Definitive Guide}, 1st edn.
\newblock O'Reilly Media, Inc. (2009)

\bibitem{ashraf}
Xiang, J., Guo, C., Aboulnaga, A.: Scalable maximum clique computation using
  mapreduce.
\newblock In: Data Engineering (ICDE), 2013 IEEE 29th International Conference
  on, pp. 74--85 (2013)

\end{thebibliography}

\label{lastpage}

\end{document}